%% file: main.tex
\documentclass[aps,prl,floatfix,twocolumn,superscriptaddress]{revtex4-2}
\input{preamble.tex}

\usepackage{bm}
\usepackage{comment}
\usepackage{float} 
\usepackage[version=4]{mhchem} 

\graphicspath{{Figures/}}

\begin{document}
\title{Optical Tweezer Arrays of Erbium Atoms}
\input{authors}


\begin{abstract}
We present the first successful trapping of single erbium atoms in an array of optical tweezers. Using a single narrow-line optical transition, we achieve deep cooling for direct tweezer loading, pairwise ejection, and continous imaging without additional recoil suppression techniques. Our tweezer wavelength choice enables us to reach the magic trapping condition by tuning the ellipticity of the trapping light. Additionally, we implement an ultrafast high-fidelity fluorescence imaging scheme using a broad transition, allowing time-resolved study of the tweezer population dynamics from many to single atoms during light-assisted collisions. In particular, we extract a pair-ejection rate that qualitatively agrees with the semiclassical predictions by the Gallagher-Pritchard model. This work represents a promising starting point for the exploration of erbium as a powerful resource for quantum simulation in optical tweezers.

\end{abstract}

\maketitle

In recent years, there has been rapid progress towards the establishment of scalable many-body quantum platforms. These platforms allow comprehensive control over both the external and internal degrees of freedom of each constituent particle, as well as their mutual interactions\,\cite{Altman2021qsa}. Powerful architectures include the use of neutral atoms\,\cite{Schaefer2020tfq,Gross2021qgm}, ions\,\cite{Monroe2021pqs_}, and molecules\,\cite{Langen2023qsm}, in which the regime of single-atom control is achieved using tailored potentials and high-resolution imaging systems. 

Among the various realizations, neutral atoms trapped in optical tweezer arrays are gaining significant momentum as a means to create quantum simulators assembled atom-by-atom\,\cite{Browaeys2020mbp,Kaufman2021qsw}. So far alkali atoms have been the preferred choice for such platforms due to their simple single-valence-electron spectrum, which facilitates the creation of these arrays\,\cite{Schlosser2001spl,Beugnon2007tdt,Nogrette2014sat,Lee2016tdr,Barredo2016aab,Endres2016aba,Bluvstein2024lqp_,Manetsch2024ata}. However, this simplicity also presents  limitations, as it restricts the range of available properties and tools available for control and manipulation. Currently, the field is expanding to more complex particles such as the two-valence-electron alkaline-earth atoms (AEAs)\,\cite{Cooper2018aea,Norcia2018mca,Finkelstein2024uqo,Cao2024mqg_,Hoelzl2024llc}, the AEA-like Yb\,\cite{Saskin2019nlc,Ma2023hfg,Norcia2024iao_}, and even diatomic\,\cite{Anderegg2019aot,Zhang2022aot,Holland2023bio} and  polyatomic molecules\,\cite{Vilas2024aot}, whose increased electronic complexity opens up unprecedented research directions. 

By moving from two- to many-valence-electron atoms, as seen in open-shell lanthanides, we reach a situation where the opportunities offered by AEAs are not only preserved, but significantly amplified. This includes an even richer range of optical transitions\,\cite{Ban2005lct} suitable for e.\,g.\,reaching the desired regime of deep laser cooling in shallow traps, isolated core excitations,  narrow-line or ultrafast imaging\,\cite{Picard2019dla,Su2024fsa}, as well as simple two-photon pathways for creating Rydberg states with large orbital momentum\,\cite{Trautmann2021sor,Kruckenhauser2023hds}.  Moreover, these atoms exhibit a strongly optically active ionic core\,\cite{Norcia2021dia} and an anisotropic character in their interaction with light due to their submerged $4f$ shell\,\cite{Dzuba2011dpa,Becher2018apo,Chalopin2018als,Bloch2024apo}. Significant progress has recently been made in the establishment of a lanthanide atom-by-atom assembler with the trapping of single dysprosium atoms in tweezer arrays\,\cite{Bloch2023tai}, while erbium, for which the Rydberg series has been recently mapped out\,\cite{Trautmann2021sor}, has yet to be loaded, trapped, and manipulated in optical tweezers.

In this work, we demonstrate trapping of individual bosonic $^{166}$Er atoms in a linear tweezer array. Similar to alkali atoms, we use a single optical transition for laser cooling, pairwise ejection, and imaging. However, unlike alkali atoms, this transition has a narrow-line character, allowing us to directly achieve deep enough cooling for efficient tweezer loading at shallow trap depths and continuous quasi-non-destructive imaging without the need to implement additional techniques to suppress recoil heating. For narrow lines, the scattering rate in a two-level cycling transition is sensitive to the differential light shift caused by the tweezer light. At our choice of tweezer wavelength, the magic condition can be reached by a simple tuning of the light ellipticity. We observe a loading peak and good imaging efficiencies at the magic condition, where this differential shift is nullified. Finally, we implement an ultrafast high-fidelity imaging scheme by flashing blue light on a microsecond time scale to drive a broad transition. This technique lets us to disentangle in-trap dynamics from imaging, allowing to study the evolution of the tweezer population under light-assisted collision pulse, including pairwise and single-atom ejection, with high resolution in time. Due to its simplicity, high level of controllability, and broad opportunities, we identify this novel platform as a highly promising resource for many quantum science goals.

\begin{figure}
	\centering
	\includegraphics[width=\columnwidth]{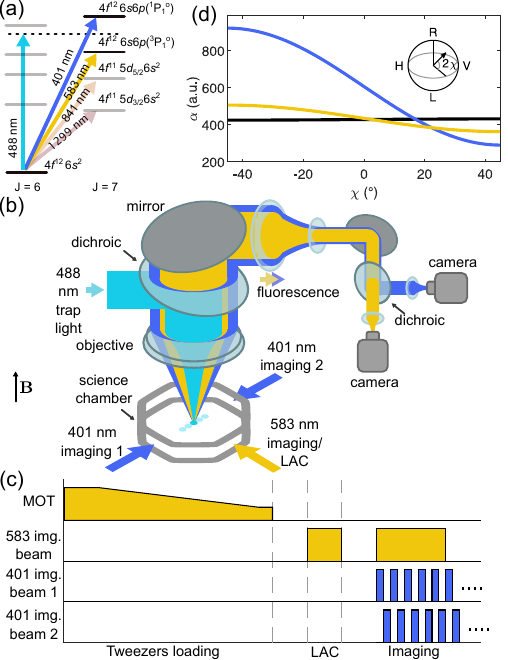}
	\caption{Optical properties of erbium, experimental setup and sequence. (a) Energy states and relevant optical transitions. (b) Schematic of the science chamber and the optical setup for trapping and imaging. (c) Typical experimental sequence; see text. The tweezer light is kept on for the whole duration of each experimental sequence. (d) Calculated $\alpha$ as a function of $\chi$ for fixed $\theta_p = \pi/2$ and $\theta_k = \pi$ and $\lambda_t=\SI{488}{\nano\meter}$. The parameter $\chi$ is defined as depicted in the inset in the Poincaré sphere. Magic conditions occur when the curves of an excited (blue and yellow line) and the ground state (black line) cross each other.}
	\label{fig:setup}
\end{figure}

Figure\,\ref{fig:setup}(a) presents the relevant erbium optical transitions from the  $m_J=-6$ ground state\,\cite{Ban2005lct}. These include the broad 401-nm [$\Gamma/(2\pi)=\SI{29.7}{\mega\hertz}$] blue line, which we use for transversal cooling, Zeeman-slowing and ultrafast imaging, and the narrow-line yellow transition at \SI{583}{\nano\meter} [$\Gamma/(2\pi)=\SI{186}{\kilo\hertz}$], employed for our five-beam MOT\,\cite{Frisch2012narrowline}, light-assisted collisions, and non-destructive imaging. We also show the 841-nm [$\Gamma/(2\pi)=\SI{8}{\kilo\hertz}$] and the clock-like 1299-nm [$\Gamma/(2\pi)=\SI{0.9}{\hertz}$] transition. The former is a promising candidate for resolved sideband and Sisyphus cooling in tweezers\,\cite{Cooper2018aea}, the latter for shelving and quantum computation\,\cite{Madjarov2020hfe, Patscheider2021ooa}.

Our experimental approach to single-atom tweezers consists of three main steps: tweezer loading from the MOT, pair-wise ejection of atoms via light-assisted collisions (LAC), and imaging. The first two are performed using the yellow narrow-line transition, whereas to probe the atoms we have the option to use either yellow or blue fluorescence, as illustrated in Fig.\,\ref{fig:setup}(b-c). For tweezer loading, we first prepare a laser-cooled spin-polarized ($m_J=-6$) cloud in a five-beam narrow-line MOT\,\cite{Frisch2012narrowline,Ilzhoefer2018tsf,supmat}. We then load the tweezer array by overlapping the MOT cloud with it for \SI{20}{\milli\second}, before switching off the MOT. For this work, we use a simple linear array of 5 optical tweezers, generated from laser light at $\lambda_t=\SI{488}{\nano\meter}$, passing through an acousto-optic deflector. We then focus the beams onto the atoms using a custom objective with a numerical aperture of NA$=0.45$ and an effective focal length of \SI{65}{\milli\meter}, creating microtraps each with a waist of about \SI{1}{\micro\meter} and a power of \SI{2.5}{\milli\watt}. The low MOT temperature of about \SI{10}{\micro\kelvin} allows for a direct trapping of atoms into comparatively shallow optical tweezers (trap depth of about \SI{150}{\micro\kelvin}) without the need for additional cooling. At the end of the loading phase, we end up with a few atoms in each trap, necessitating a dedicated stage of LAC, as discussed later. Finally, we perform fluorescence imaging of the trapped sample and collect the scattered photons through the aforementioned objective. Remarkably, erbium offers the possibility to obtain fluorescence images in two different regimes. The first is based on the yellow light for continuous non-destructive imaging\,\cite{Aikawa:2012,Miranda2015sri}. The second relies on the strong blue transition, which allows to collect blue fluorescence at \si{\micro\second}-short exposure times\,\cite{Picard2019dla}, recently observed in lattice-confined erbium systems\,\cite{Su2023dqs_}. To compensate for the photon recoil during absorption, we use two counter-propagating blue beams. Additionally, to avoid unwanted interference and other broadening effects\,\cite{Su2023dqs_} between the beams, we activate them in an alternating fashion, resulting in a \SI{30}{\micro\second}-long train of pulses, each lasting \SI{2.5}{\micro\second}; see Fig.\,\ref{fig:setup}(c).

In tweezer experiments, a crucial effect arises when trapped atoms are exposed to near-resonant light, such as during loading, cooling, and imaging. The tweezer light induces a position-dependent quadratic AC-Stark shift of the bare energy levels\,\cite{Kaufman2021qsw}. The strength of the light shift depends on the dynamic polarizability $\alpha$ of the atoms and varies for each atomic level. When the shift gets comparable to the linewidth of the near-resonant excited state of the transition, it often becomes essential to reach the so-called magic condition, where the corresponding differential light shift is zero. In alkali species, where only the scalar polarizability is normally significant, the magic condition is quite restrictive as it requires a specific value of $\lambda_t$ for each two-level system. In contrast, anisotropic atomic species like erbium offer much greater flexibility due to $\alpha$ being a complex anisotropic tensor\,\cite{Lepers2014aot, Becher2018apo}. Here, $\alpha$ additionally depends on several tweezer-light parameters: the angle $\theta_k$ ($\theta_p$) between the light-propagation direction $\mathbf{\hat{e}}_k$ (polarization vector $\mathbf{u}$)  and the quantization axis, and the ellipticity parameter $\chi$, where $i\sin(2\chi) = (\mathbf{u}^* \times \mathbf{u})\cdot\mathbf{\hat{e}}_k$\,\cite{Rosenband2018sel}. For $\lambda_t=\SI{488}{\nano\meter}$, our calculations show magic conditions can be found by varying $\chi$ for both the blue and yellow cycling transitions; see Fig.\,\ref{fig:setup}(d).

\begin{figure}
	\centering
	\includegraphics[scale=1]{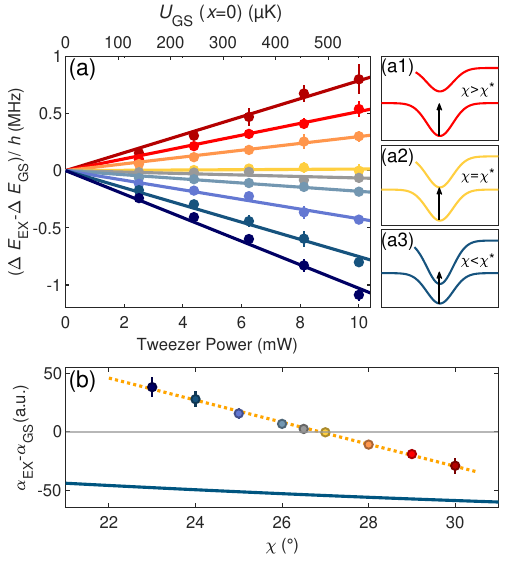}
	\caption {Polarizability of erbium and magic condition (a) Measured AC-Stark shift of the $\SI{583}{\nano\meter}$ transition as a function of the trapping light power for various $\chi$. Solid lines are linear fits to the data, with the intercept fixed to zero. The images on the right show the radial dependence of the ground and excited levels when the resonance has a positive (a1), zero (a2) and negative (a3) shift. (b) Differential polarizability, obtained from the slopes of the linear fits in (a), as a function of $\chi$. The solid line is a theoretical calculation, the gray line indicates the magic condition, and the dotted line is a linear fit.}
	\label{fig:polarizability} 
\end{figure}

\begin{figure}
	\centering
	\includegraphics[width=\columnwidth]{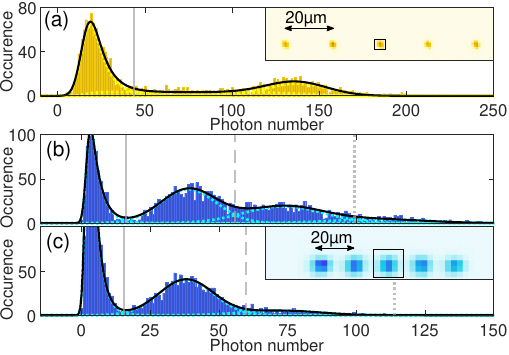}
    \caption{Narrow-line and ultrafast imaging. (a) Histogram obtained from 2000 repetitions with the narrow-line yellow imaging on the central trap of the 5-site array with $t_\text{LAC}=\SI{70}{\milli\second}$, ${\Delta_\text{LAC}=-1.4\,\Gamma}$, and $I =  0.4\,I_{\rm sat}$. For the imaging, we illuminate the atoms during $\SI{70}{\milli\second}$ with $I = 0.25\,I_{\rm sat}$ and $\Delta = -1.2\,\Gamma$.  (b-c) Histograms obtained with the ultrafast blue imaging for (b) $t_\text{LAC}=\SI{30}{\milli\second}$ and (c) $t_\text{LAC}=\SI{70}{\milli\second}$. Here, we use imaging parameters $I \approx 5\,I_{\rm sat}$ and $\Delta = 0$. In (a-c), the solid black line shows the result of the multi-peak fit, while the dotted lines show separate contributions from noise and atoms. The vertical lines indicate the classification thresholds. The insets shows the average image of the arrays, where the different size of the point-spread-function is related to the atom diffusion during imaging. }
    \label{fig:histo}
\end{figure}

We experimentally determine the magic condition for the yellow transition via fluorescence spectroscopy. After loading the tweezer and switching off the MOT, we set the desired value of $\chi$ within $\SI{300}{\milli\second}$ using a pair of motorized waveplates. We then shine a single horizontal yellow probe beam of intensity $I = 0.4\,I_{\rm sat}$ on the trapped atoms for $\SI{1}{\milli\second}$. When scanning the probe frequency, we find a peak in the number of collected fluorescence photons.
By repeating the measurement at different tweezer powers, we observe the expected linear shift of the peak, whose slope directly gives the differential light shift, as shown in Fig.\,\ref{fig:polarizability}(a).  
By taking similar scans for different values of $\chi$, we are able to identify the magic condition -- where the light shift is zero -- close to $\chi^*=\SI{26.86(5)}{\degree}$, as shown in Fig.\,\ref{fig:polarizability}(b). Note that we observe a substantial mismatch between the theoretically predicted and the experimentally observed magic $\chi$. Potential reasons for the mismatch include the inherent difficulty in calculating $\alpha$ for lanthanides, uncertainties in determining the tweezer beam waist and ellipticity, and effects related to the breakdown of the paraxial approximation that may emerge for tightly focused beams\,\cite{Thompson2013car}. Notably, similar mismatches have recently been observed with dysprosium\,\cite{Bloch2024apo}, suggesting that further dedicated investigations may be necessary. Henceforth the measurements are performed with optical tweezers in the experimentally determined magic condition. 

\begin{figure}
	\centering
	\includegraphics[width=\columnwidth]{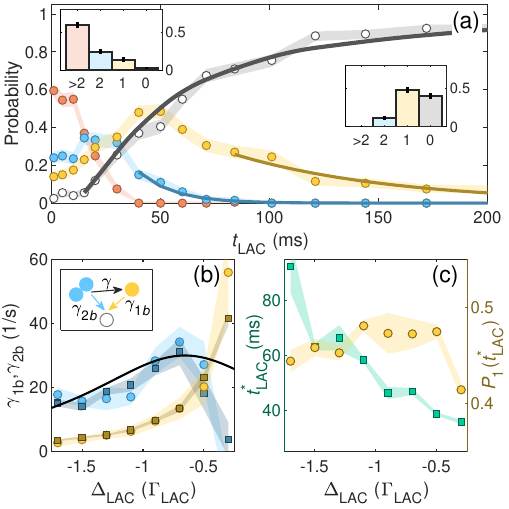}
    \caption{Evolution of tweezer population as a function of LAC pulse duration and detuning, at fixed $I = 0.4\,I_{\rm sat}$. (a) Probability of having an empty trap (gray), single atom (yellow), two atoms (blue) and more than two atoms (red) as a function of $t_\text{LAC}$ at $\Delta_\text{LAC}=-0.7\,\Gamma$. Each data point is derived from 200 runs. Solid lines indicate exponential fits to the one and two atom probabilities and the result of the rate equation fit for the empty trap. The insets show the histograms of the atom number probability at $t_\text{LAC}=\SI{1}{\milli\second}$ and $t_\text{LAC}=\SI{50}{\milli\second}$. (b) Extracted decay rates $\gamma_{1b}$ (yellow) and $\gamma_{2b}$ (blue) from the exponential fits (circles) and from the rate equation fits (squares). The solid line depicts the (rescaled) Gallagher-Pritchard collision rate. (c) Extracted LAC time $t^*_\text{LAC}$ (squares) where the probability for one atom is maximal, together with the corresponding probability $P_1(t_\text{LAC}^*)$ (circles) as a function of $\Delta_\text{LAC}$. (a-c) Shaded areas indicate the 1-$\sigma$ confidence interval.}
	\label{fig:losses}
\end{figure}

As previously mentioned, to achieve single-atom occupancy, we apply a dedicated stage of LAC, triggering pairwise ejection\,\cite{Weiner1999eat,Schlosser2002cbi,Fung2016sap}. For this, we illuminate the system with a single near-resonant beam of yellow light for a duration $t_\text{LAC}$ of tens of milliseconds. As also observed in AEAs\,\cite{Cooper2018aea, Norcia2018mca,Saskin2019nlc} and dysprosium\,\cite{Bloch2023tai}, such an additional stage is necessary because the regime of collisional blockade, common in alkali, is not fully reached when operating with narrow-line MOTs and \SI{}{\micro\meter}-waist tweezers\,\cite{Schlosser2002cbi}. After $t_\text{LAC}$, we perform fluorescence imaging and collect scattered photons over a given exposure time. We then construct a cumulative histogram from thousands of repetitions to probe the atom-number occupancy.

In a first set of experiments, we use yellow fluorescence imaging for probing. The narrow-line character of the yellow transition allows for long exposure times during which the trapped atoms undergo continuous absorption-emission cycles with minimal loss due to recoil heating. From the histogram shown in Fig.\,\ref{fig:histo}(a), we observe two distinct peaks: a zero-atom background peak and a single-atom peak \,\cite{supmat}. The absence of a two-atom signal indicates successful LAC-induced pair ejection, whereas the flat photon distribution, bridging the zero- and single-atom peaks, suggests that a small fraction of single atoms is lost before the end of the imaging process\,\cite{Cooper2018aea,Bloch2023tai}. The area of the bridge depends on the imaging exposure time, which, for the measurements of Fig.\,\ref{fig:histo}(a), is purposefully chosen to be long (\SI{70}{\milli\second}) to highlight this effect. We analyze the histogram using a multi-peak fitting function from which we extract the relevant parameters, including the peak positions, amplitudes, and the height of the loss bridge\,\cite{supmat}. The optimal threshold position to categorize the number of atoms present in each image is retrieved by minimizing the estimated wrong classifications given by the overlap between the peaks. This method also maximizes the fidelity $F_{0,1}$, which quantifies the accuracy in classifying zero or one atom\,\cite{Cooper2018aea,Norcia2018mca,supmat}. For the parameters of Fig.\,\ref{fig:histo}(a), we detect a single atom with a probability of $P_1= 43(1)\%$ and a fidelity of $F_{0,1}=0.94(1)$. One-body losses responsible for the bridge and the $P_1 <50\%$ might be due to recoil heating during LAC and imaging \cite{Cooper2018aea}, which could be compensated in future experiments with additional cooling stages, or to two-photon transitions driven by the yellow and tweezer light\,\cite{Bloch2023tai}.

In a second set of experiments, we use the ultrafast blue imaging; see Fig.\,\ref{fig:setup}. Figure\,\ref{fig:histo}(b-c) show examples of histograms for (b) $t_\text{LAC}=\SI{30}{\milli\second}$ and (c) \SI{70}{\milli\second}. The first histogram demonstrates the capability to image and resolve multiple atoms. To separate the contributions from one, two and more atoms, we extract the corresponding thresholds from the multi-peak fit. These will be used later to convert the number of photons from a single image into the number of atoms. Here, we obtain a single-atom classification fidelity of $F_1=0.91(1)$\,\cite{supmat} and a probability of successfully loading a single atom of $P_1 = 41(1)\%$, with a residual chance $P_{\ge2}=21(1)\%$ of loading more than one atom. By increasing the duration of the LAC pulse we can progressively decrease the probability of loading multiple atoms; see e.\,g.\,Fig.\,\ref{fig:histo}(c) for $t_\text{LAC} = \SI{70}{\milli\second}$.

The ultrafast blue imaging allows us to disregard processes occurring during detection, enabling a study of the atom-number occupation dynamics during exposure to yellow light with high temporal resolution. This capability allows to trace the relative contributions of single- and pairwise ejections, which are crucial for single-atom-tweezer operations. Using the aforementioned categorization protocol based on a blue histogram, Fig.\,\ref{fig:losses}(a) shows the occupation probabilities for zero, one, two, and more than two atoms as a function of $t_\text{LAC}$ for a detuning $\Delta_\text{LAC}=-0.7\,\Gamma$. Initially, we observe the expected high probability of having multiple atoms, which decreases over time. As this happens, the probabilities for single and double atom occupations increase. The double-atom one reaches its maximum first and then gradually decreases, while the single-atom probability continues to rise, peaking at a later time $t^*_\text{LAC}$. After their respective maxima, both the single and double atom probabilities exhibit exponential tails, which for the former arises mainly from the one-body recoil heating whereas the latter additionally undergoes pair-ejection losses. 

To quantify the strength of both processes, we extract the one- and two-body loss rates, $\gamma_{1b}$ and $\gamma_{2b}$ using two different techniques. In the first, we simply fit an exponential decay to the tails of $P_1$ and $P_2$, whereas for the second, we fit a rate-equation model\,\cite{Sortais2012spa} to $P_0$, as detailed in Ref.\,\cite{supmat}. Both methods agree to each other within the error bars. Additionally, we repeat the above measurement and the analysis as a function of $\Delta_\text{LAC}$. As shown in Fig.\,\ref{fig:losses}(b), we see that $\gamma_{2b}$ dominates over $\gamma_{1b}$ at large detuning. When approaching the resonance, the pair-ejection rate reaches its maximum and rapidly decreases. Remarkably, this behavior is qualitatively well reproduced by the Gallagher-Pritchard model\,\cite{Gallagher1989eco}; see solid line. The rate $\gamma_{1b}$ instead undergoes a rapid increase while approaching resonance. The inversion of the strength of the two rates suggest the existence of an optimal detuning.  Interestingly, while the optimal LAC duration $t^*_\text{LAC}$ strongly depends on the detuning, we observe that the maximum one-body occupancy $P_1(t^*_\text{LAC})$ is close to the stochastic $50\,\%$ limit for a large range of $\Delta_\text{LAC}$, as shown in Fig.\,\ref{fig:losses}(c). This is a very promising starting condition for implementing reconfigurable tweezer arrays.

In this work, we realized a new platform for quantum simulation and quantum information processing by demonstrating trapping and imaging of single erbium atoms in a tweezer array. We are able to reach magic conditions for our narrow 583-nm transition at our tweezer trapping wavelength by employing the anisotropic polarizability present for lanthanides, and we characterized the population dynamics during the LAC pulse using ultrafast imaging. The combination of tweezers with ultrafast number-resolved imaging will allow us to investigate the interplay between LAC, different cooling/imaging schemes and other loss processes in an unique way. Additionally, the rich optical spectrum of erbium opens the door to many fascinating possibilities, ranging from trappable Rydberg atoms and direct excitation routes to high angular momentum states, to direct imaging of Rydberg atoms.


\begin{acknowledgments}
We thank Julián Maloberti and Amal El Arrach for contributions in the early stage of the experiment. We thank Riccardo Donofrio for contributions in data collection. We thank Antoine Browaeys, Igor Ferrier-Barbut, Matthew Norcia for valuable discussions. We also thank the other members of the Dipolar Quantum Gases group at the University of Innsbruck for useful discussions.
This work was supported by the European Research Council through the Advanced Grant DyMETEr (No.\,101054500), the QuantERA grant MAQS by the Austrian Science Fund FWF (No.\,I4391-N), a NextGeneration EU Grant AQuSIM through the Austrian Research Promotion Agency (FFG) (No.\,FO999896041), and the Austrian Science Fund (FWF) Cluster of Excellence QuantA (\href{https://doi.org/10.55776/COE1}{10.55776/COE1}). A.\,O. is supported by the Swiss National Science Foundation through the Postdoc.Mobility fellowship P500PT\texttt{\char`_}211074. M.\,L. acknowledges support from the NeoDip project (ANR-19-CE30-0018-01 from ‘Agence Nationale de la Recherche’). We also acknowledge the Innsbruck Laser Core Facility, financed by the Austrian Federal Ministry of Science, Research and Economy. 
\end{acknowledgments}

\bibliography{reference_final}


\clearpage
\include{supmat.tex}

\end{document}

%% file: preamble.tex
\usepackage{graphicx,
            caption,
            subcaption,
            amsmath,
            braket,
            ragged2e,
            xcolor,
            xspace,
            nicefrac,
            lipsum,
            nameref,
            textcomp,
            urwchancal,
            float,
            upgreek,
            isotope,
            hyperref,
            cleveref,
            siunitx,
            placeins, 
            scrextend, 
            url}
\DeclareCaptionLabelSeparator{dot}{. }
\makeatletter
\def\justified{
	\let\\\@normalcr
	\@rightskip\z@skip \rightskip\@rightskip
	\leftskip\z@skip
	\parindent 0em\relax
	\setlength{\parfillskip}{0pt plus 1fil}}
\DeclareCaptionJustification{justified}{\justified}
\hypersetup{colorlinks=true, linkcolor=blue,citecolor=blue,urlcolor=blue}
\def\unit #1 #2 {\SI{#1}{#2}\xspace}
\def\range #1 #2 #3 {\SIrange{#1}{#2}{#3}\xspace}
\DeclareSIUnit\gauss{G}

\newcommand{\myref}[2][]{Fig.~\hyperref[#2]{\ref*{#2}#1}}
\newcommand{\Myref}[2][]{Figure~\hyperref[#2]{\ref*{#2}#1}}
\newcommand{\Mytabref}[2][]{Table~\hyperref[#2]{\ref*{#2}#1}}

\newcommand{\uibk}[0]{\affiliation{Universität Innsbruck, Institut für Experimentalphysik, Technikerstraße 25, 6020 Innsbruck, Austria}}
\newcommand{\iqoqi}[0]{\affiliation{ Institut für Quantenoptik und Quanteninformation, Österreichische Akademie der Wissenschaften, Technikerstraße 21a, 6020 Innsbruck, Austria}}

%% file: authors.tex
\date{\today}

 
\author{D.\,S.\,Grün}
\thanks{These authors contributed equally to this work.}
\iqoqi
\uibk

\author{S.\,J.\,M.\,White}
\thanks{These authors contributed equally to this work.}
\iqoqi
\uibk

\author{A.\,Ortu}
\iqoqi
\uibk

\author{A.\,Di Carli}
\iqoqi
\uibk

\author{H.\,Edri}
\iqoqi
\uibk

\author{M.\,Lepers}
\affiliation{Laboratoire Interdisciplinaire Carnot de Bourgogne, CNRS, Université de Bourgogne, 21078 Dijon, France}

\author{M.\,J.\,Mark}
\iqoqi
\uibk

\author{F.\,Ferlaino}
\thanks{Correspondence should be addressed to \mbox{\url{Francesca.Ferlaino@uibk.ac.at}}}
\iqoqi
\uibk

%% file: supmat.tex
\setcounter{equation}{0}
\setcounter{figure}{0}
\setcounter{table}{0}
\setcounter{page}{1}
\makeatletter
\renewcommand{\theequation}{S\arabic{equation}}
\renewcommand{\thefigure}{S\arabic{figure}}


\section*{Supplemental Material}
\subsection{Experimental Sequence}
\label{sec:supmat_setup}

Our vacuum setup is similar to that described in Ref.\,\cite{Ilzhoefer2018tsf}. In short, a high temperature oven (CreaTec Dual Filament Cell) emits a beam of erbium vapour through an aperture into the transversal cooling chamber, where the atomic beam is cooled transversely using two retro-reflected \SI{401}{\nano\metre} beams slightly red detuned from the atomic resonance. It continues into the Zeeman slower, where the longitudinal velocity of the atoms is decreased to the capture velocity of the narrow-line MOT using again the broad \SI{401}{\nano\metre} transition. The slowed atoms are captured in the main chamber by a narrow-line five-beam MOT operating on the intercombination line at \SI{583}{\nano\metre}. After $\SI{50}{\milli\second}$ of MOT loading, 
we smoothly compress the MOT (cMOT) in $\simeq\SI{230}{\milli\second}$. After the compression, the cMOT cloud, containing $\sim 10^6$ spin-polarized ($m_J=-6$) atoms at about \SI{10}{\micro\kelvin}, spatially overlaps with the tweezer array. 
The five-beam geometry allows the direct placement of a high numerical aperture objective (Special Optics) with NA$=0.45$ directly above the chamber. This objective is used for both tweezer generation and fluorescence imaging. Additionally, though not used in this work, an array of eight in-vacuum electrodes for electric field control, two multi-channel plate detectors (Hamamatsu F4655-10S184) for the simultaneous detection of ions and electrons, and two circular and two dipole antennas are located within the chamber. The electrodes slightly reduce the available numerical aperture to NA$=0.42$.

\subsection{Tweezer generation}

Our tweezer light at \SI{488}{\nano\metre} (generated by an AzurLight Systems ALS-BL-488-2-E-CP-SF) is split into several beams using an acousto-optic deflector (AA Optoelectronics DTS-400-488), driven with a multi-tone signal generated by an arbitrary waveform generator (Spectrum Instrumentation M4i6631-x8). The resultant beams are expanded to a diameter of about $\SI{40}{\milli\metre}$ before being reflected into the objective using a custom 4" dichroic mirror (LENS-Optics) designed to be reflective in a small neighbourhood around \SI{488}{\nano\metre} and highly transmissive at our imaging wavelengths of \SI{401}{\nano\metre} and \SI{583}{\nano\metre}. We typically load the traps with a power of \SI{2.5}{\milli\watt} per tweezer. Assuming the ground state polarizability at \SI{488}{\nano\metre} to be $\alpha_{\rm GS}=430\,$a.u., we measure a waist of about \SI{1}{\micro\metre} from parametric heating measurements, which translates into a calculated trap depth of $U_0/k_B \sim\SI{150}{\micro\kelvin}$.

\subsection{Anisotropic polarizability}
\label{sec:supmat_an}
The total polarizability can be  decomposed in the sum of three contributions proportional to the scalar, vectorial, and tensorial components represented as $\alpha_s$, $\alpha_v$ and $\alpha_t$ respectively~\cite{Li2017aot,Becher2018apo}:
\begin{multline}
 	\alpha(\lambda_t) = \alpha_s(\lambda_t)  -i\left(\mathbf{u}^*\times\mathbf{u}  \right)\cdot\mathbf{\hat{e}}_k\cos\theta_k \frac{m_J}{2J} \alpha_v(\lambda_t) \\
  +\frac{3 m_J^2-J(J+1)}{J(2J-1)} \frac{3\cos^2\theta_p-1}{2} \alpha_t(\lambda_t).
	\label{eq:polarizability}
\end{multline}
The angles $\theta_k$ and $\theta_p$ are between the quantization axis and the propagation direction $\mathbf{\hat{e}}_k$, and the polarization of the light, respectively. The dependence on $\mathbf{u}$ can be replaced by the ellipticity parameter $\chi$, given by ${(\mathbf{u}^*\times\mathbf{u})\cdot\mathbf{\hat{e}}_k = i\sin(2\chi)}$~\cite{Rosenband2018sel}. The multi-parameter dependence of polarizability offers significant flexibility in achieving the magic condition.

\begin{figure}[htb]
    \centering
    \includegraphics[width=\columnwidth]{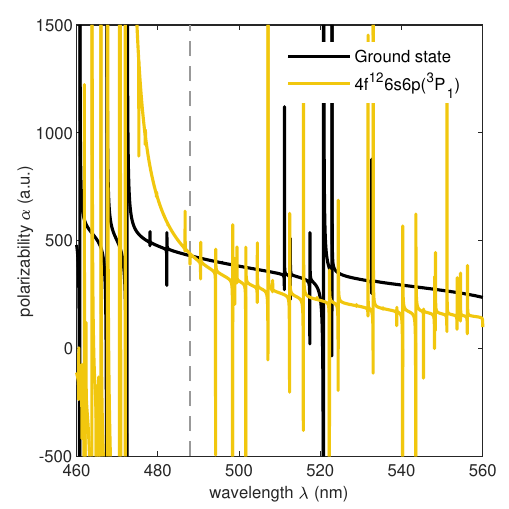}
    \caption{Calculated polarizability of the groundstate and excited state of the 583-nm transition of erbium as a function of the wavelength. Parameters used are $\theta_k=\pi$, $\theta_p=\pi/2$, and $\chi=0^\circ$. The dashed line indicates our trapping light wavelength.}
    \label{fig:polarizabilitywavelength}
\end{figure}

To calculate the polarizability of the states of interest, we numerically evaluate Eq.\,\ref{eq:polarizability} using the theoretically predicted coefficients for the scalar, vectorial and tensorial polarizabilities.
Figure\,\ref{fig:polarizabilitywavelength} plots $\alpha(\lambda)$ for the ground state and excited state of the 583-nm transition within a region of interest around the trapping wavelength $\lambda_t=\SI{488}{\nano\meter}$ with $\theta_k$ and $\theta_p$, resembling our experimental configuration, and an ellipticity angle $\chi=0^\circ$.

Here, we can identify -- beside many narrow resonant features -- the slowly varying background behavior which resemble the magic condition close to our chosen trap wavelength. Additional tuning of $\chi$ allows us to fine tune the relative polarizability, as shown in Fig.\,\ref{fig:setup}(d) in the main article.

\subsection{Polarizability measurement procedure}

\begin{figure}[htb]
	\centering
	\includegraphics[scale=1]{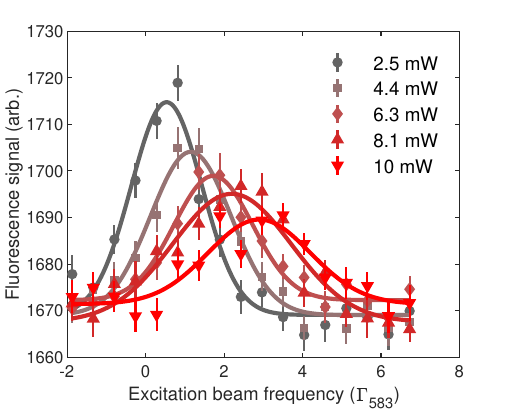}
    \caption{Polarizability measurement. Series of fluorescence spectra of erbium taken at various trap powers at ellipticity value $\chi=29^\circ$. Solid lines denote Gaussian fits to the data with which we extract the peak positions.}
    \label{fig:light-shift}
\end{figure}

For the measurement of the differential polarizability presented in Fig.\,\ref{fig:polarizability} in the main article, we perform fluorescence spectroscopy of the atoms while trapped in the tweezers. For this, the tweezer traps are loaded with many atoms at a fixed power and ellipticity of the trapping light to ensure stable starting conditions. Afterwards, the power and ellipticity of the trapping light are ramped to the final values within a few hundreds of ms (limited by the rotation speed of the motorized waveplates) and then illuminated by our 583-nm imaging beam for \SI{1}{\milli\second}. We record the captured fluorescence signal as a function of the detuning from resonance for various trap powers, see Fig.\,\ref{fig:light-shift} for an example of such a measurement series. From these data we extract the relative AC Stark shift induced by the trapping light as the shift in peak position, derived from a Gaussian fit to the data, at different powers.

We calibrate the ellipticity of the trap light by measuring its polarization after it passes through the main chamber using a commercial polarimeter (ThorLabs PAX1000VIS/M) for different configurations of the motorised $\lambda/4$ and $\lambda/2$ waveplates. Note that we observe a minor dependence of the measured light shift on the azimuth angle of the polarization ellipse. This can be explained by experimental imperfections e.g. an angle deviation between the $B$-field and the tweezer light propagation direction, and a possible breakdown of the paraxial approximation in the focus of the objective. We therefore use a range of waveplate configurations where we are able to vary the elliptical component of the polarisation with a minimal variation of the azimuthal component.

\subsection{Fluorescence imaging}

We perform fluorescence imaging by collecting scattered photons from the atoms. For the continuous imaging at \SI{583}{\nano\metre}, we use a single imaging beam with horizontal polarization and an inclination of $10^\circ$ to the horizontal plane. In case of the fast imaging at \SI{401}{\nano\metre}, we implemented a pair of two counter-propagating beams in the horizontal plane both having horizontal polarization. During imaging, we ensure that the applied magnetic field along the vertical direction is sufficiently strong to isolate the $\sigma^-$ transition. This configuration maximises the fluorescence signal collected through the objective due to the resulting dipole emission pattern.

After separating the emitted fluorescence signal from the tweezer light via the dichroic mirror, we guide the light through a system of achromatic lenses and reflective angle-tunable filters (Semrock TLP01-628 and TSP01-628). Those filters form a bandpass filter around \SI{583}{\nano\metre} with a calculated full width at half maximum (FWHM) of less than \SI{4}{\nano\metre} while reflecting more than \SI{99}{\percent} of the light at \SI{401}{\nano\metre}. The separated wavelengths are then focused onto the cameras. For blue imaging we always use an EMCCD camera (Andor iXon Ultra 897). For yellow imaging, we normally use a sCMOS camera (Andor Zyla 4.2P), apart from Fig.\,\ref{fig:histo}(a) in the main text which was also taken with the EMCCD. We place bandpass filters with a FWHM of \SI{10}{\nano\metre} over the apertures of both cameras (ThorLabs FBH400-10 and FBH580-10 for the iXon and Zyla camera respectively) in order to reduce background light.

\subsection{Image processing}

Every histogram or e.g. datapoint in Fig.\,\ref{fig:losses}(a) of the main article is derived from one measurement containing from several hundreds up to a few thousands single shot images. First, we calculate the average image of the measurement and extract the rough positions of each tweezer by fitting Gaussians to the integrated densities. Next, for each trap, we define a region-of-interest (ROI) whose position is optimized further by maximizing the integrated signal strength within. We optimized the size of the ROI with respect to maximizing the signal while minimizing the background noise, and set it to $5\times5$ pixels for both imaging schemes.

Finally, we extract the integrated signal strength $N_\text{counts}$ per image and trap by summing up the counts of the corresponding ROI, and derive the detected photon number $n$ via

\begin{equation*}
    n = \frac{N_\text{counts}-N_\text{offset}}{C_\text{cam}}
\end{equation*}

with $N_\text{offset}$ a fixed offset added by the camera and $C_\text{cam}$ a conversion factor, taking into account the camera's quantum efficiency, the EM-gain and the conversion factor of electrons to counts, giving ${C_\text{cam}=34.53\,}$counts/photon for 401-nm imaging and ${C_\text{cam}=43.17\,}$counts/photon for 583-nm imaging. We cross-check that the number of detected photons is consistent within our experimental uncertainties with the expected number of photons scattered from the atoms when taking into account the finite solid angle of the collected light (NA$=0.42$), losses due to absorption/reflection on the optical path to the camera, the detuning and intensity of the imaging light as well as the radiation emission pattern of the atoms.

\subsection{Histogram analysis}
\label{sec:supmat_an}

In this work, we present histograms of the accumulated photon number $n$ per trap extracted from many experimental repetitions. We fit the envelope of the histogram with a function composed of several terms: First, the background noise peak is modelled by a convolution of a Gaussian with amplitude $a_0$, mean $\mu_0$ and standard deviation $\sigma_0$, with an exponential tail of constant $c$:

\begin{equation*}
f_0(n) = a_0\frac{c}{2}\,e^{\frac{c}{2}(2\mu_0+c\sigma_0^2-2n)}\,\text{erf}\left(\frac{\mu_0+c\sigma_0^2-n}{\sqrt{2}\sigma_0}\right).
\end{equation*}

The signal of each atom gives a Poissonian distribution, broadened by a Gaussian due to the camera readout noise.
We model the first peak with a Gaussian with amplitude $a_1$, mean $\mu_1$ and standard deviation $\sigma_1$

\begin{equation*}
    f_1(n) = a_1\,e^{-\frac{1}{2}[(n-\mu_1)/\sigma_1]^2}.
\end{equation*}

For a peak corresponding to a higher number of atoms $k$, we still consider a Gaussian shape, but with mean and width which are functions of the noise and first peak parameters, while its amplitude $a_k$ is left free. Namely, the mean will be a multiple of the single-atom mean signal, after taking into account the noise median $\mu_0+\log2/c$ and the Poissonian broadening $\sqrt{k}$.

\begin{multline*}
    f_k(n) = \\
    a_k\,\exp\left\{{-\frac{1}{2}\left[\frac{n-k\mu_1+(k-1)(\mu_0+\log2/c)}{\sigma_1 \sqrt{k}}\right]^2}\right\}.    
\end{multline*}

For the 583-nm imaging histograms, we additionally model a bridge of loss by integrating the single-atom Gaussian over a mean $\mu_1$ that varies homogeneously between the 1-atom peak and the noise peak~\cite{Bloch2023tai},

\begin{equation*}
f_b(n) = a_b\,\text{erf}\left(\frac{n-\mu_1}{\sqrt{2}\sigma_1}\right)\,\text{erf}\left(\frac{\mu_0+c\sigma_0^2-n}{\sqrt{2}\sigma_0}\right).
\end{equation*}

The final fitting function for the whole histogram is given by

\begin{equation*}
f(n) = f_0(n)+f_b(n)+\sum_{k=1}^{k_\text{max}} f_k(n),
\label{eq:supmat_fit}
\end{equation*}

where $k_\text{max}$ is the maximum number of atoms we expect to be present.

To identify the number of atoms from the number of detected photons for individual images, we extract thresholds between the peaks corresponding to the noise, 1-atom, and 2-atom signals. Each threshold $x_k$, distinguishing between $k-1$ and $k$ atoms, is found by maximizing the fidelity

\begin{equation*}
    F_{k-1,k} = \frac{1}{C}\max_{x_k}\left[ \int_{-\infty}^{x_k} \sum_{j=0}^{k-1} f_j(n)\, dn + \int_{x_k}^\infty \sum_{j=k}^{k_\text{max}} f_j(n)\, dn \right].
\end{equation*}

with $C=\int_{-\infty}^\infty \sum_{k=0}^{k_\text{max}} f_k(n) \,dn$ the normalization constant. For the threshold between noise and one atom $x_1$ in case of 583-nm imaging we additionally include the loss bridge function. Note that e.g. $F_{0,1}$ gives the fidelity to have a correct classification between an occupied and an empty tweezer. In general, the fidelity of correctly classifying $N$ atoms is given by the ratio of correct classifications to all classifications within the corresponding threshold interval:

\begin{equation*}
    F_N = \frac{\int_{x_{N}}^{x_{N+1}} f_{N}(n) \,dn}{\sum_{j=0}^{k_\text{max}} \int_{x_{N}}^{x_{N+1}}  f_j(n) \,dn} .
\end{equation*}

\begin{figure}
\includegraphics[width=\columnwidth]{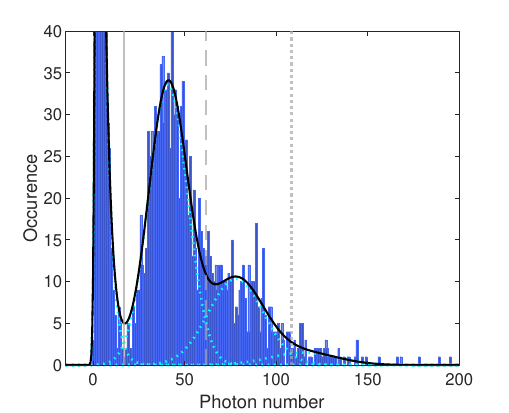}
\caption{Reference histogram for the thresholds used to calculate the occupation probabilities in Fig.\,\ref{fig:losses} in the article. Bright blue dashed lines show the single fitted components of the noise and the signal of 1 and 2 atomic occupation number. Together, they result in the fitted black curve.
The gray vertical lines show the thresholds $x_1=17$ photons (solid), $x_2=61$ photons (dashed) and $x_3=108$ photons (dotted).
}
\label{fig:supmat_histo}
\end{figure}

For the analysis of Fig.\,\ref{fig:losses} in the main article we use fixed thresholds for all values of $t_\text{LAC}$ and $\Delta_\text{LAC}$.
The thresholds are obtained from a histogram consisting of about 2000 images collected for $t_\text{LAC}=\SI{45}{\milli\second}$ and $\Delta_\text{LAC}=-\Gamma$, shown in Fig.\,\ref{fig:supmat_histo}. This particular $t_\text{LAC}$ was chosen so to be at the beginning of the range used for the fits. $\Delta_\text{LAC}$ was chosen instead to be at the center of the range of detuning values we used.

For each experimental run we extract the detected atom number according to those thresholds and derive the probability of detecting $k$ atoms $P_k$ as the ratio of experimental runs with $k$ atoms to the total number of runs for the same conditions. The error of this probability is estimated by the 68\,\% binomial proportion confidence interval.

\subsection{Loss rate extraction}
\label{sec:supmat_rate}

The analysis of the loss rates in Fig.\,\ref{fig:losses} in the main article is done with two independent fitting processes. First, we directly fit the long-time behavior of the one-atom probability $P_1$ and the two-atom probability $P_2$ using exponential decay equations to get rates $\gamma_\text{1b}$ and $\gamma$. For fitting $P_1$, we include data from the time where the probability to have two atoms falls below $P_2<0.02$, while for fitting $P_2$, we include only data where $P_{\geq2}<0.05$. This ensures that the dynamics do not include growth terms from higher atom states. The loss rate $\gamma$ out of the two-body state should be a combination of two-body and one-body processes. We therefore derive $\gamma_\text{2b}=\gamma-2\gamma_\text{1b}$, assuming independent loss for each single atom present.

Second, we derived a rate equation that describes the direct increase of the zero-atom signal coming from the two-atom state via light-assisted collisions with rate $\gamma_\text{2b}$ and from the one-atom state with rate $\gamma_\text{1b}$:

\begin{equation*}
    \dot{P_0} = \gamma_\text{2b}\cdot P_2 + \gamma_\text{1b}\cdot P_1
\end{equation*}

We numerically integrate this rate equation using the experimentally measured single and two-atom probabilities $P_1$ and $P_2$, and use $\gamma_\text{1b}$ and $\gamma_\text{2b}$ as the fitting parameters for a fit to the experimental data in $P_0$. We start the fit from the time where the probability to have more than two atoms falls below $P_{\geq2}<0.5$, as our error in classification reduces with the mean atom number.

%% file: main.bbl
\providecommand{\noopsort}[1]{}\providecommand{\singleletter}[1]{#1}%
\begin{thebibliography}{55}%
\makeatletter
\providecommand \@ifxundefined [1]{%
 \@ifx{#1\undefined}
}%
\providecommand \@ifnum [1]{%
 \ifnum #1\expandafter \@firstoftwo
 \else \expandafter \@secondoftwo
 \fi
}%
\providecommand \@ifx [1]{%
 \ifx #1\expandafter \@firstoftwo
 \else \expandafter \@secondoftwo
 \fi
}%
\providecommand \natexlab [1]{#1}%
\providecommand \enquote  [1]{``#1''}%
\providecommand \bibnamefont  [1]{#1}%
\providecommand \bibfnamefont [1]{#1}%
\providecommand \citenamefont [1]{#1}%
\providecommand \href@noop [0]{\@secondoftwo}%
\providecommand \href [0]{\begingroup \@sanitize@url \@href}%
\providecommand \@href[1]{\@@startlink{#1}\@@href}%
\providecommand \@@href[1]{\endgroup#1\@@endlink}%
\providecommand \@sanitize@url [0]{\catcode `\\12\catcode `\$12\catcode
  `\&12\catcode `\#12\catcode `\^12\catcode `\_12\catcode `\%12\relax}%
\providecommand \@@startlink[1]{}%
\providecommand \@@endlink[0]{}%
\providecommand \url  [0]{\begingroup\@sanitize@url \@url }%
\providecommand \@url [1]{\endgroup\@href {#1}{\urlprefix }}%
\providecommand \urlprefix  [0]{URL }%
\providecommand \Eprint [0]{\href }%
\providecommand \doibase [0]{https://doi.org/}%
\providecommand \selectlanguage [0]{\@gobble}%
\providecommand \bibinfo  [0]{\@secondoftwo}%
\providecommand \bibfield  [0]{\@secondoftwo}%
\providecommand \translation [1]{[#1]}%
\providecommand \BibitemOpen [0]{}%
\providecommand \bibitemStop [0]{}%
\providecommand \bibitemNoStop [0]{.\EOS\space}%
\providecommand \EOS [0]{\spacefactor3000\relax}%
\providecommand \BibitemShut  [1]{\csname bibitem#1\endcsname}%
\let\auto@bib@innerbib\@empty
\bibitem [{\citenamefont {Altman}\ \emph {et~al.}(2021)\citenamefont {Altman},
  \citenamefont {Brown}, \citenamefont {Carleo}, \citenamefont {Carr},
  \citenamefont {Demler}, \citenamefont {Chin}, \citenamefont {DeMarco},
  \citenamefont {Economou}, \citenamefont {Eriksson}, \citenamefont {Fu} \emph
  {et~al.}}]{Altman2021qsa}%
  \BibitemOpen
  \bibfield  {author} {\bibinfo {author} {\bibfnamefont {E.}~\bibnamefont
  {Altman}}, \bibinfo {author} {\bibfnamefont {K.~R.}\ \bibnamefont {Brown}},
  \bibinfo {author} {\bibfnamefont {G.}~\bibnamefont {Carleo}}, \bibinfo
  {author} {\bibfnamefont {L.~D.}\ \bibnamefont {Carr}}, \bibinfo {author}
  {\bibfnamefont {E.}~\bibnamefont {Demler}}, \bibinfo {author} {\bibfnamefont
  {C.}~\bibnamefont {Chin}}, \bibinfo {author} {\bibfnamefont {B.}~\bibnamefont
  {DeMarco}}, \bibinfo {author} {\bibfnamefont {S.~E.}\ \bibnamefont
  {Economou}}, \bibinfo {author} {\bibfnamefont {M.~A.}\ \bibnamefont
  {Eriksson}}, \bibinfo {author} {\bibfnamefont {K.-M.~C.}\ \bibnamefont {Fu}},
  \emph {et~al.},\ }\bibfield  {title} {\bibinfo {title} {Quantum simulators:
  Architectures and opportunities},\ }\href
  {https://doi.org/10.1103/PRXQuantum.2.017003} {\bibfield  {journal} {\bibinfo
   {journal} {PRX Quantum}\ }\textbf {\bibinfo {volume} {2}},\ \bibinfo {pages}
  {017003} (\bibinfo {year} {2021})}\BibitemShut {NoStop}%
\bibitem [{\citenamefont {Sch{\"a}fer}\ \emph {et~al.}(2020)\citenamefont
  {Sch{\"a}fer}, \citenamefont {Fukuhara}, \citenamefont {Sugawa},
  \citenamefont {Takasu},\ and\ \citenamefont {Takahashi}}]{Schaefer2020tfq}%
  \BibitemOpen
  \bibfield  {author} {\bibinfo {author} {\bibfnamefont {F.}~\bibnamefont
  {Sch{\"a}fer}}, \bibinfo {author} {\bibfnamefont {T.}~\bibnamefont
  {Fukuhara}}, \bibinfo {author} {\bibfnamefont {S.}~\bibnamefont {Sugawa}},
  \bibinfo {author} {\bibfnamefont {Y.}~\bibnamefont {Takasu}},\ and\ \bibinfo
  {author} {\bibfnamefont {Y.}~\bibnamefont {Takahashi}},\ }\bibfield  {title}
  {\bibinfo {title} {{Tools for quantum simulation with ultracold atoms in
  optical lattices}},\ }\href {https://doi.org/10.1038/s42254-020-0195-3}
  {\bibfield  {journal} {\bibinfo  {journal} {Nat. Rev. Phys.}\ }\textbf
  {\bibinfo {volume} {2}},\ \bibinfo {pages} {411} (\bibinfo {year}
  {2020})}\BibitemShut {NoStop}%
\bibitem [{\citenamefont {Gross}\ and\ \citenamefont
  {Bakr}(2021)}]{Gross2021qgm}%
  \BibitemOpen
  \bibfield  {author} {\bibinfo {author} {\bibfnamefont {C.}~\bibnamefont
  {Gross}}\ and\ \bibinfo {author} {\bibfnamefont {W.~S.}\ \bibnamefont
  {Bakr}},\ }\bibfield  {title} {\bibinfo {title} {Quantum gas microscopy for
  single atom and spin detection},\ }\href
  {https://doi.org/10.1038/s41567-021-01370-5} {\bibfield  {journal} {\bibinfo
  {journal} {Nature Physics}\ }\textbf {\bibinfo {volume} {17}},\ \bibinfo
  {pages} {1316} (\bibinfo {year} {2021})}\BibitemShut {NoStop}%
\bibitem [{\citenamefont {Monroe}\ \emph {et~al.}(2021)\citenamefont {Monroe},
  \citenamefont {Campbell}, \citenamefont {Duan}, \citenamefont {Gong},
  \citenamefont {Gorshkov}, \citenamefont {Hess}, \citenamefont {Islam},
  \citenamefont {Kim}, \citenamefont {Linke}, \citenamefont {Pagano} \emph
  {et~al.}}]{Monroe2021pqs_}%
  \BibitemOpen
  \bibfield  {author} {\bibinfo {author} {\bibfnamefont {C.}~\bibnamefont
  {Monroe}}, \bibinfo {author} {\bibfnamefont {W.~C.}\ \bibnamefont
  {Campbell}}, \bibinfo {author} {\bibfnamefont {L.-M.}\ \bibnamefont {Duan}},
  \bibinfo {author} {\bibfnamefont {Z.-X.}\ \bibnamefont {Gong}}, \bibinfo
  {author} {\bibfnamefont {A.~V.}\ \bibnamefont {Gorshkov}}, \bibinfo {author}
  {\bibfnamefont {P.~W.}\ \bibnamefont {Hess}}, \bibinfo {author}
  {\bibfnamefont {R.}~\bibnamefont {Islam}}, \bibinfo {author} {\bibfnamefont
  {K.}~\bibnamefont {Kim}}, \bibinfo {author} {\bibfnamefont {N.~M.}\
  \bibnamefont {Linke}}, \bibinfo {author} {\bibfnamefont {G.}~\bibnamefont
  {Pagano}}, \emph {et~al.},\ }\bibfield  {title} {\bibinfo {title}
  {Programmable quantum simulations of spin systems with trapped ions},\ }\href
  {https://doi.org/10.1103/RevModPhys.93.025001} {\bibfield  {journal}
  {\bibinfo  {journal} {Rev. Mod. Phys.}\ }\textbf {\bibinfo {volume} {93}},\
  \bibinfo {pages} {025001} (\bibinfo {year} {2021})}\BibitemShut {NoStop}%
\bibitem [{\citenamefont {Langen}\ \emph {et~al.}(2023)\citenamefont {Langen},
  \citenamefont {Valtolina}, \citenamefont {Wang},\ and\ \citenamefont
  {Ye}}]{Langen2023qsm}%
  \BibitemOpen
  \bibfield  {author} {\bibinfo {author} {\bibfnamefont {T.}~\bibnamefont
  {Langen}}, \bibinfo {author} {\bibfnamefont {G.}~\bibnamefont {Valtolina}},
  \bibinfo {author} {\bibfnamefont {D.}~\bibnamefont {Wang}},\ and\ \bibinfo
  {author} {\bibfnamefont {J.}~\bibnamefont {Ye}},\ }\bibfield  {title}
  {\bibinfo {title} {Quantum state manipulation and science of ultracold
  molecules},\ }\href@noop {} {\  (\bibinfo {year} {2023})},\ \Eprint
  {https://arxiv.org/abs/2305.13445} {arXiv:2305.13445 [cond-mat.quant-gas]}
  \BibitemShut {NoStop}%
\bibitem [{\citenamefont {Browaeys}\ and\ \citenamefont
  {Lahaye}(2020)}]{Browaeys2020mbp}%
  \BibitemOpen
  \bibfield  {author} {\bibinfo {author} {\bibfnamefont {A.}~\bibnamefont
  {Browaeys}}\ and\ \bibinfo {author} {\bibfnamefont {T.}~\bibnamefont
  {Lahaye}},\ }\bibfield  {title} {\bibinfo {title} {Many-body physics with
  individually controlled {R}ydberg atoms},\ }\href
  {https://doi.org/10.1038/s41567-019-0733-z} {\bibfield  {journal} {\bibinfo
  {journal} {Nature Physics}\ }\textbf {\bibinfo {volume} {16}},\ \bibinfo
  {pages} {132} (\bibinfo {year} {2020})}\BibitemShut {NoStop}%
\bibitem [{\citenamefont {Kaufman}\ and\ \citenamefont
  {Ni}(2021)}]{Kaufman2021qsw}%
  \BibitemOpen
  \bibfield  {author} {\bibinfo {author} {\bibfnamefont {A.~M.}\ \bibnamefont
  {Kaufman}}\ and\ \bibinfo {author} {\bibfnamefont {K.-K.}\ \bibnamefont
  {Ni}},\ }\bibfield  {title} {\bibinfo {title} {Quantum science with optical
  tweezer arrays of ultracold atoms and molecules},\ }\href
  {https://doi.org/10.1038/s41567-021-01357-2} {\bibfield  {journal} {\bibinfo
  {journal} {Nature Physics}\ }\textbf {\bibinfo {volume} {17}},\ \bibinfo
  {pages} {1324} (\bibinfo {year} {2021})}\BibitemShut {NoStop}%
\bibitem [{\citenamefont {Schlosser}\ \emph {et~al.}(2001)\citenamefont
  {Schlosser}, \citenamefont {Reymond}, \citenamefont {Protsenko},\ and\
  \citenamefont {Grangier}}]{Schlosser2001spl}%
  \BibitemOpen
  \bibfield  {author} {\bibinfo {author} {\bibfnamefont {N.}~\bibnamefont
  {Schlosser}}, \bibinfo {author} {\bibfnamefont {G.}~\bibnamefont {Reymond}},
  \bibinfo {author} {\bibfnamefont {I.}~\bibnamefont {Protsenko}},\ and\
  \bibinfo {author} {\bibfnamefont {P.}~\bibnamefont {Grangier}},\ }\bibfield
  {title} {\bibinfo {title} {Sub-{P}oissonian loading of single atoms in a
  microscopic dipole trap},\ }\href {https://doi.org/10.1038/35082512}
  {\bibfield  {journal} {\bibinfo  {journal} {Nature}\ }\textbf {\bibinfo
  {volume} {411}},\ \bibinfo {pages} {1024} (\bibinfo {year}
  {2001})}\BibitemShut {NoStop}%
\bibitem [{\citenamefont {Beugnon}\ \emph {et~al.}(2007)\citenamefont
  {Beugnon}, \citenamefont {Tuchendler}, \citenamefont {Marion}, \citenamefont
  {Ga{\"e}tan}, \citenamefont {Miroshnychenko}, \citenamefont {Sortais},
  \citenamefont {Lance}, \citenamefont {Jones}, \citenamefont {Messin},
  \citenamefont {Browaeys},\ and\ \citenamefont {Grangier}}]{Beugnon2007tdt}%
  \BibitemOpen
  \bibfield  {author} {\bibinfo {author} {\bibfnamefont {J.}~\bibnamefont
  {Beugnon}}, \bibinfo {author} {\bibfnamefont {C.}~\bibnamefont {Tuchendler}},
  \bibinfo {author} {\bibfnamefont {H.}~\bibnamefont {Marion}}, \bibinfo
  {author} {\bibfnamefont {A.}~\bibnamefont {Ga{\"e}tan}}, \bibinfo {author}
  {\bibfnamefont {Y.}~\bibnamefont {Miroshnychenko}}, \bibinfo {author}
  {\bibfnamefont {Y.~R.~P.}\ \bibnamefont {Sortais}}, \bibinfo {author}
  {\bibfnamefont {A.~M.}\ \bibnamefont {Lance}}, \bibinfo {author}
  {\bibfnamefont {M.~P.~A.}\ \bibnamefont {Jones}}, \bibinfo {author}
  {\bibfnamefont {G.}~\bibnamefont {Messin}}, \bibinfo {author} {\bibfnamefont
  {A.}~\bibnamefont {Browaeys}},\ and\ \bibinfo {author} {\bibfnamefont
  {P.}~\bibnamefont {Grangier}},\ }\bibfield  {title} {\bibinfo {title}
  {Two-dimensional transport and transfer of a single atomic qubit in optical
  tweezers},\ }\href {https://doi.org/10.1038/nphys698} {\bibfield  {journal}
  {\bibinfo  {journal} {Nature Physics}\ }\textbf {\bibinfo {volume} {3}},\
  \bibinfo {pages} {696} (\bibinfo {year} {2007})}\BibitemShut {NoStop}%
\bibitem [{\citenamefont {Nogrette}\ \emph {et~al.}(2014)\citenamefont
  {Nogrette}, \citenamefont {Labuhn}, \citenamefont {Ravets}, \citenamefont
  {Barredo}, \citenamefont {B\'eguin}, \citenamefont {Vernier}, \citenamefont
  {Lahaye},\ and\ \citenamefont {Browaeys}}]{Nogrette2014sat}%
  \BibitemOpen
  \bibfield  {author} {\bibinfo {author} {\bibfnamefont {F.}~\bibnamefont
  {Nogrette}}, \bibinfo {author} {\bibfnamefont {H.}~\bibnamefont {Labuhn}},
  \bibinfo {author} {\bibfnamefont {S.}~\bibnamefont {Ravets}}, \bibinfo
  {author} {\bibfnamefont {D.}~\bibnamefont {Barredo}}, \bibinfo {author}
  {\bibfnamefont {L.}~\bibnamefont {B\'eguin}}, \bibinfo {author}
  {\bibfnamefont {A.}~\bibnamefont {Vernier}}, \bibinfo {author} {\bibfnamefont
  {T.}~\bibnamefont {Lahaye}},\ and\ \bibinfo {author} {\bibfnamefont
  {A.}~\bibnamefont {Browaeys}},\ }\bibfield  {title} {\bibinfo {title}
  {Single-atom trapping in holographic 2d arrays of microtraps with arbitrary
  geometries},\ }\href {https://doi.org/10.1103/PhysRevX.4.021034} {\bibfield
  {journal} {\bibinfo  {journal} {Phys. Rev. X}\ }\textbf {\bibinfo {volume}
  {4}},\ \bibinfo {pages} {021034} (\bibinfo {year} {2014})}\BibitemShut
  {NoStop}%
\bibitem [{\citenamefont {Lee}\ \emph {et~al.}(2016)\citenamefont {Lee},
  \citenamefont {Kim},\ and\ \citenamefont {Ahn}}]{Lee2016tdr}%
  \BibitemOpen
  \bibfield  {author} {\bibinfo {author} {\bibfnamefont {W.}~\bibnamefont
  {Lee}}, \bibinfo {author} {\bibfnamefont {H.}~\bibnamefont {Kim}},\ and\
  \bibinfo {author} {\bibfnamefont {J.}~\bibnamefont {Ahn}},\ }\bibfield
  {title} {\bibinfo {title} {Three-dimensional rearrangement of single atoms
  using actively controlled optical microtraps},\ }\href
  {https://doi.org/10.1364/OE.24.009816} {\bibfield  {journal} {\bibinfo
  {journal} {Opt. Express}\ }\textbf {\bibinfo {volume} {24}},\ \bibinfo
  {pages} {9816} (\bibinfo {year} {2016})}\BibitemShut {NoStop}%
\bibitem [{\citenamefont {Barredo}\ \emph {et~al.}(2016)\citenamefont
  {Barredo}, \citenamefont {de~L{\'e}s{\'e}leuc}, \citenamefont {Lienhard},
  \citenamefont {Lahaye},\ and\ \citenamefont {Browaeys}}]{Barredo2016aab}%
  \BibitemOpen
  \bibfield  {author} {\bibinfo {author} {\bibfnamefont {D.}~\bibnamefont
  {Barredo}}, \bibinfo {author} {\bibfnamefont {S.}~\bibnamefont
  {de~L{\'e}s{\'e}leuc}}, \bibinfo {author} {\bibfnamefont {V.}~\bibnamefont
  {Lienhard}}, \bibinfo {author} {\bibfnamefont {T.}~\bibnamefont {Lahaye}},\
  and\ \bibinfo {author} {\bibfnamefont {A.}~\bibnamefont {Browaeys}},\
  }\bibfield  {title} {\bibinfo {title} {{An atom-by-atom assembler of
  defect-free arbitrary two-dimensional atomic arrays}},\ }\href
  {https://doi.org/10.1126/science.aah3778} {\bibfield  {journal} {\bibinfo
  {journal} {Science}\ }\textbf {\bibinfo {volume} {354}},\ \bibinfo {pages}
  {1021} (\bibinfo {year} {2016})}\BibitemShut {NoStop}%
\bibitem [{\citenamefont {Endres}\ \emph {et~al.}(2016)\citenamefont {Endres},
  \citenamefont {Bernien}, \citenamefont {Keesling}, \citenamefont {Levine},
  \citenamefont {Anschuetz}, \citenamefont {Krajenbrink}, \citenamefont
  {Senko}, \citenamefont {Vuleti{\'c}}, \citenamefont {Greiner},\ and\
  \citenamefont {Lukin}}]{Endres2016aba}%
  \BibitemOpen
  \bibfield  {author} {\bibinfo {author} {\bibfnamefont {M.}~\bibnamefont
  {Endres}}, \bibinfo {author} {\bibfnamefont {H.}~\bibnamefont {Bernien}},
  \bibinfo {author} {\bibfnamefont {A.}~\bibnamefont {Keesling}}, \bibinfo
  {author} {\bibfnamefont {H.}~\bibnamefont {Levine}}, \bibinfo {author}
  {\bibfnamefont {E.~R.}\ \bibnamefont {Anschuetz}}, \bibinfo {author}
  {\bibfnamefont {A.}~\bibnamefont {Krajenbrink}}, \bibinfo {author}
  {\bibfnamefont {C.}~\bibnamefont {Senko}}, \bibinfo {author} {\bibfnamefont
  {V.}~\bibnamefont {Vuleti{\'c}}}, \bibinfo {author} {\bibfnamefont
  {M.}~\bibnamefont {Greiner}},\ and\ \bibinfo {author} {\bibfnamefont {M.~D.}\
  \bibnamefont {Lukin}},\ }\bibfield  {title} {\bibinfo {title} {{Atom-by-atom
  assembly of defect-free one-dimensional cold atom arrays}},\ }\href
  {https://doi.org/10.1126/science.aah3752} {\bibfield  {journal} {\bibinfo
  {journal} {Science}\ }\textbf {\bibinfo {volume} {354}},\ \bibinfo {pages}
  {1024} (\bibinfo {year} {2016})}\BibitemShut {NoStop}%
\bibitem [{\citenamefont {Bluvstein}\ \emph {et~al.}(2024)\citenamefont
  {Bluvstein}, \citenamefont {Evered}, \citenamefont {Geim}, \citenamefont
  {Li}, \citenamefont {Zhou}, \citenamefont {Manovitz}, \citenamefont {Ebadi},
  \citenamefont {Cain}, \citenamefont {Kalinowski}, \citenamefont {Hangleiter}
  \emph {et~al.}}]{Bluvstein2024lqp_}%
  \BibitemOpen
  \bibfield  {author} {\bibinfo {author} {\bibfnamefont {D.}~\bibnamefont
  {Bluvstein}}, \bibinfo {author} {\bibfnamefont {S.~J.}\ \bibnamefont
  {Evered}}, \bibinfo {author} {\bibfnamefont {A.~A.}\ \bibnamefont {Geim}},
  \bibinfo {author} {\bibfnamefont {S.~H.}\ \bibnamefont {Li}}, \bibinfo
  {author} {\bibfnamefont {H.}~\bibnamefont {Zhou}}, \bibinfo {author}
  {\bibfnamefont {T.}~\bibnamefont {Manovitz}}, \bibinfo {author}
  {\bibfnamefont {S.}~\bibnamefont {Ebadi}}, \bibinfo {author} {\bibfnamefont
  {M.}~\bibnamefont {Cain}}, \bibinfo {author} {\bibfnamefont {M.}~\bibnamefont
  {Kalinowski}}, \bibinfo {author} {\bibfnamefont {D.}~\bibnamefont
  {Hangleiter}}, \emph {et~al.},\ }\bibfield  {title} {\bibinfo {title}
  {Logical quantum processor based on reconfigurable atom arrays},\ }\href
  {https://doi.org/10.1038/s41586-023-06927-3} {\bibfield  {journal} {\bibinfo
  {journal} {Nature}\ }\textbf {\bibinfo {volume} {626}},\ \bibinfo {pages}
  {58} (\bibinfo {year} {2024})}\BibitemShut {NoStop}%
\bibitem [{\citenamefont {Manetsch}\ \emph {et~al.}(2024)\citenamefont
  {Manetsch}, \citenamefont {Nomura}, \citenamefont {Bataille}, \citenamefont
  {Leung}, \citenamefont {Lv},\ and\ \citenamefont {Endres}}]{Manetsch2024ata}%
  \BibitemOpen
  \bibfield  {author} {\bibinfo {author} {\bibfnamefont {H.~J.}\ \bibnamefont
  {Manetsch}}, \bibinfo {author} {\bibfnamefont {G.}~\bibnamefont {Nomura}},
  \bibinfo {author} {\bibfnamefont {E.}~\bibnamefont {Bataille}}, \bibinfo
  {author} {\bibfnamefont {K.~H.}\ \bibnamefont {Leung}}, \bibinfo {author}
  {\bibfnamefont {X.}~\bibnamefont {Lv}},\ and\ \bibinfo {author}
  {\bibfnamefont {M.}~\bibnamefont {Endres}},\ }\bibfield  {title} {\bibinfo
  {title} {A tweezer array with 6100 highly coherent atomic qubits},\
  }\href@noop {} {\  (\bibinfo {year} {2024})},\ \Eprint
  {https://arxiv.org/abs/2403.12021} {arXiv:2403.12021 [quant-ph]} \BibitemShut
  {NoStop}%
\bibitem [{\citenamefont {Cooper}\ \emph {et~al.}(2018)\citenamefont {Cooper},
  \citenamefont {Covey}, \citenamefont {Madjarov}, \citenamefont {Porsev},
  \citenamefont {Safronova},\ and\ \citenamefont {Endres}}]{Cooper2018aea}%
  \BibitemOpen
  \bibfield  {author} {\bibinfo {author} {\bibfnamefont {A.}~\bibnamefont
  {Cooper}}, \bibinfo {author} {\bibfnamefont {J.~P.}\ \bibnamefont {Covey}},
  \bibinfo {author} {\bibfnamefont {I.~S.}\ \bibnamefont {Madjarov}}, \bibinfo
  {author} {\bibfnamefont {S.~G.}\ \bibnamefont {Porsev}}, \bibinfo {author}
  {\bibfnamefont {M.~S.}\ \bibnamefont {Safronova}},\ and\ \bibinfo {author}
  {\bibfnamefont {M.}~\bibnamefont {Endres}},\ }\bibfield  {title} {\bibinfo
  {title} {Alkaline-earth atoms in optical tweezers},\ }\href
  {https://doi.org/10.1103/PhysRevX.8.041055} {\bibfield  {journal} {\bibinfo
  {journal} {Phys. Rev. X}\ }\textbf {\bibinfo {volume} {8}},\ \bibinfo {pages}
  {041055} (\bibinfo {year} {2018})}\BibitemShut {NoStop}%
\bibitem [{\citenamefont {Norcia}\ \emph {et~al.}(2018)\citenamefont {Norcia},
  \citenamefont {Young},\ and\ \citenamefont {Kaufman}}]{Norcia2018mca}%
  \BibitemOpen
  \bibfield  {author} {\bibinfo {author} {\bibfnamefont {M.~A.}\ \bibnamefont
  {Norcia}}, \bibinfo {author} {\bibfnamefont {A.~W.}\ \bibnamefont {Young}},\
  and\ \bibinfo {author} {\bibfnamefont {A.~M.}\ \bibnamefont {Kaufman}},\
  }\bibfield  {title} {\bibinfo {title} {Microscopic control and detection of
  ultracold strontium in optical-tweezer arrays},\ }\href
  {https://doi.org/10.1103/PhysRevX.8.041054} {\bibfield  {journal} {\bibinfo
  {journal} {Phys. Rev. X}\ }\textbf {\bibinfo {volume} {8}},\ \bibinfo {pages}
  {041054} (\bibinfo {year} {2018})}\BibitemShut {NoStop}%
\bibitem [{\citenamefont {Finkelstein}\ \emph {et~al.}(2024)\citenamefont
  {Finkelstein}, \citenamefont {Tsai}, \citenamefont {Sun}, \citenamefont
  {Scholl}, \citenamefont {Direkci}, \citenamefont {Gefen}, \citenamefont
  {Choi}, \citenamefont {Shaw},\ and\ \citenamefont
  {Endres}}]{Finkelstein2024uqo}%
  \BibitemOpen
  \bibfield  {author} {\bibinfo {author} {\bibfnamefont {R.}~\bibnamefont
  {Finkelstein}}, \bibinfo {author} {\bibfnamefont {R.~B.-S.}\ \bibnamefont
  {Tsai}}, \bibinfo {author} {\bibfnamefont {X.}~\bibnamefont {Sun}}, \bibinfo
  {author} {\bibfnamefont {P.}~\bibnamefont {Scholl}}, \bibinfo {author}
  {\bibfnamefont {S.}~\bibnamefont {Direkci}}, \bibinfo {author} {\bibfnamefont
  {T.}~\bibnamefont {Gefen}}, \bibinfo {author} {\bibfnamefont
  {J.}~\bibnamefont {Choi}}, \bibinfo {author} {\bibfnamefont {A.~L.}\
  \bibnamefont {Shaw}},\ and\ \bibinfo {author} {\bibfnamefont
  {M.}~\bibnamefont {Endres}},\ }\bibfield  {title} {\bibinfo {title}
  {Universal quantum operations and ancilla-based readout for tweezer clocks},\
  }\href@noop {} {\  (\bibinfo {year} {2024})},\ \Eprint
  {https://arxiv.org/abs/2402.16220} {arXiv:2402.16220 [quant-ph]} \BibitemShut
  {NoStop}%
\bibitem [{\citenamefont {Cao}\ \emph {et~al.}(2024)\citenamefont {Cao},
  \citenamefont {Eckner}, \citenamefont {Yelin}, \citenamefont {Young},
  \citenamefont {Jandura}, \citenamefont {Yan}, \citenamefont {Kim},
  \citenamefont {Pupillo}, \citenamefont {Ye}, \citenamefont {Oppong} \emph
  {et~al.}}]{Cao2024mqg_}%
  \BibitemOpen
  \bibfield  {author} {\bibinfo {author} {\bibfnamefont {A.}~\bibnamefont
  {Cao}}, \bibinfo {author} {\bibfnamefont {W.~J.}\ \bibnamefont {Eckner}},
  \bibinfo {author} {\bibfnamefont {T.~L.}\ \bibnamefont {Yelin}}, \bibinfo
  {author} {\bibfnamefont {A.~W.}\ \bibnamefont {Young}}, \bibinfo {author}
  {\bibfnamefont {S.}~\bibnamefont {Jandura}}, \bibinfo {author} {\bibfnamefont
  {L.}~\bibnamefont {Yan}}, \bibinfo {author} {\bibfnamefont {K.}~\bibnamefont
  {Kim}}, \bibinfo {author} {\bibfnamefont {G.}~\bibnamefont {Pupillo}},
  \bibinfo {author} {\bibfnamefont {J.}~\bibnamefont {Ye}}, \bibinfo {author}
  {\bibfnamefont {N.~D.}\ \bibnamefont {Oppong}}, \emph {et~al.},\ }\bibfield
  {title} {\bibinfo {title} {Multi-qubit gates and '{S}chr\"odinger cat' states
  in an optical clock},\ }\href@noop {} {\  (\bibinfo {year} {2024})},\ \Eprint
  {https://arxiv.org/abs/2402.16289} {arXiv:2402.16289 [quant-ph]} \BibitemShut
  {NoStop}%
\bibitem [{\citenamefont {H\"olzl}\ \emph {et~al.}(2024)\citenamefont
  {H\"olzl}, \citenamefont {G\"otzelmann}, \citenamefont {Pultinevicius},
  \citenamefont {Wirth},\ and\ \citenamefont {Meinert}}]{Hoelzl2024llc}%
  \BibitemOpen
  \bibfield  {author} {\bibinfo {author} {\bibfnamefont {C.}~\bibnamefont
  {H\"olzl}}, \bibinfo {author} {\bibfnamefont {A.}~\bibnamefont
  {G\"otzelmann}}, \bibinfo {author} {\bibfnamefont {E.}~\bibnamefont
  {Pultinevicius}}, \bibinfo {author} {\bibfnamefont {M.}~\bibnamefont
  {Wirth}},\ and\ \bibinfo {author} {\bibfnamefont {F.}~\bibnamefont
  {Meinert}},\ }\bibfield  {title} {\bibinfo {title} {Long-lived circular
  {R}ydberg qubits of alkaline-earth atoms in optical tweezers},\ }\href
  {https://doi.org/10.1103/PhysRevX.14.021024} {\bibfield  {journal} {\bibinfo
  {journal} {Phys. Rev. X}\ }\textbf {\bibinfo {volume} {14}},\ \bibinfo
  {pages} {021024} (\bibinfo {year} {2024})}\BibitemShut {NoStop}%
\bibitem [{\citenamefont {Saskin}\ \emph {et~al.}(2019)\citenamefont {Saskin},
  \citenamefont {Wilson}, \citenamefont {Grinkemeyer},\ and\ \citenamefont
  {Thompson}}]{Saskin2019nlc}%
  \BibitemOpen
  \bibfield  {author} {\bibinfo {author} {\bibfnamefont {S.}~\bibnamefont
  {Saskin}}, \bibinfo {author} {\bibfnamefont {J.~T.}\ \bibnamefont {Wilson}},
  \bibinfo {author} {\bibfnamefont {B.}~\bibnamefont {Grinkemeyer}},\ and\
  \bibinfo {author} {\bibfnamefont {J.~D.}\ \bibnamefont {Thompson}},\
  }\bibfield  {title} {\bibinfo {title} {Narrow-line cooling and imaging of
  ytterbium atoms in an optical tweezer array},\ }\href
  {https://doi.org/10.1103/PhysRevLett.122.143002} {\bibfield  {journal}
  {\bibinfo  {journal} {Phys. Rev. Lett.}\ }\textbf {\bibinfo {volume} {122}},\
  \bibinfo {pages} {143002} (\bibinfo {year} {2019})}\BibitemShut {NoStop}%
\bibitem [{\citenamefont {Ma}\ \emph {et~al.}(2023)\citenamefont {Ma},
  \citenamefont {Liu}, \citenamefont {Peng}, \citenamefont {Zhang},
  \citenamefont {Jandura}, \citenamefont {Claes}, \citenamefont {Burgers},
  \citenamefont {Pupillo}, \citenamefont {Puri},\ and\ \citenamefont
  {Thompson}}]{Ma2023hfg}%
  \BibitemOpen
  \bibfield  {author} {\bibinfo {author} {\bibfnamefont {S.}~\bibnamefont
  {Ma}}, \bibinfo {author} {\bibfnamefont {G.}~\bibnamefont {Liu}}, \bibinfo
  {author} {\bibfnamefont {P.}~\bibnamefont {Peng}}, \bibinfo {author}
  {\bibfnamefont {B.}~\bibnamefont {Zhang}}, \bibinfo {author} {\bibfnamefont
  {S.}~\bibnamefont {Jandura}}, \bibinfo {author} {\bibfnamefont
  {J.}~\bibnamefont {Claes}}, \bibinfo {author} {\bibfnamefont {A.~P.}\
  \bibnamefont {Burgers}}, \bibinfo {author} {\bibfnamefont {G.}~\bibnamefont
  {Pupillo}}, \bibinfo {author} {\bibfnamefont {S.}~\bibnamefont {Puri}},\ and\
  \bibinfo {author} {\bibfnamefont {J.~D.}\ \bibnamefont {Thompson}},\
  }\bibfield  {title} {\bibinfo {title} {High-fidelity gates and mid-circuit
  erasure conversion in an atomic qubit},\ }\href
  {https://doi.org/10.1038/s41586-023-06438-1} {\bibfield  {journal} {\bibinfo
  {journal} {Nature}\ }\textbf {\bibinfo {volume} {622}},\ \bibinfo {pages}
  {279} (\bibinfo {year} {2023})}\BibitemShut {NoStop}%
\bibitem [{\citenamefont {Norcia}\ \emph {et~al.}(2024)\citenamefont {Norcia},
  \citenamefont {Kim}, \citenamefont {Cairncross}, \citenamefont {Stone},
  \citenamefont {Ryou}, \citenamefont {Jaffe}, \citenamefont {Brown},
  \citenamefont {Barnes}, \citenamefont {Battaglino}, \citenamefont
  {Bohdanowicz} \emph {et~al.}}]{Norcia2024iao_}%
  \BibitemOpen
  \bibfield  {author} {\bibinfo {author} {\bibfnamefont {M.~A.}\ \bibnamefont
  {Norcia}}, \bibinfo {author} {\bibfnamefont {H.}~\bibnamefont {Kim}},
  \bibinfo {author} {\bibfnamefont {W.~B.}\ \bibnamefont {Cairncross}},
  \bibinfo {author} {\bibfnamefont {M.}~\bibnamefont {Stone}}, \bibinfo
  {author} {\bibfnamefont {A.}~\bibnamefont {Ryou}}, \bibinfo {author}
  {\bibfnamefont {M.}~\bibnamefont {Jaffe}}, \bibinfo {author} {\bibfnamefont
  {M.~O.}\ \bibnamefont {Brown}}, \bibinfo {author} {\bibfnamefont
  {K.}~\bibnamefont {Barnes}}, \bibinfo {author} {\bibfnamefont
  {P.}~\bibnamefont {Battaglino}}, \bibinfo {author} {\bibfnamefont {T.~C.}\
  \bibnamefont {Bohdanowicz}}, \emph {et~al.},\ }\bibfield  {title} {\bibinfo
  {title} {Iterative assembly of $^{171}${Y}b atom arrays in cavity-enhanced
  optical lattices},\ }\href@noop {} {\  (\bibinfo {year} {2024})},\ \Eprint
  {https://arxiv.org/abs/2401.16177} {arXiv:2401.16177 [quant-ph]} \BibitemShut
  {NoStop}%
\bibitem [{\citenamefont {Anderegg}\ \emph {et~al.}(2019)\citenamefont
  {Anderegg}, \citenamefont {Cheuk}, \citenamefont {Bao}, \citenamefont
  {Burchesky}, \citenamefont {Ketterle}, \citenamefont {Ni},\ and\
  \citenamefont {Doyle}}]{Anderegg2019aot}%
  \BibitemOpen
  \bibfield  {author} {\bibinfo {author} {\bibfnamefont {L.}~\bibnamefont
  {Anderegg}}, \bibinfo {author} {\bibfnamefont {L.~W.}\ \bibnamefont {Cheuk}},
  \bibinfo {author} {\bibfnamefont {Y.}~\bibnamefont {Bao}}, \bibinfo {author}
  {\bibfnamefont {S.}~\bibnamefont {Burchesky}}, \bibinfo {author}
  {\bibfnamefont {W.}~\bibnamefont {Ketterle}}, \bibinfo {author}
  {\bibfnamefont {K.-K.}\ \bibnamefont {Ni}},\ and\ \bibinfo {author}
  {\bibfnamefont {J.~M.}\ \bibnamefont {Doyle}},\ }\bibfield  {title} {\bibinfo
  {title} {{An optical tweezer array of ultracold molecules}},\ }\href
  {https://doi.org/10.1126/science.aax1265} {\bibfield  {journal} {\bibinfo
  {journal} {Science}\ }\textbf {\bibinfo {volume} {365}},\ \bibinfo {pages}
  {1156} (\bibinfo {year} {2019})}\BibitemShut {NoStop}%
\bibitem [{\citenamefont {Zhang}\ \emph {et~al.}(2022)\citenamefont {Zhang},
  \citenamefont {Picard}, \citenamefont {Cairncross}, \citenamefont {Wang},
  \citenamefont {Yu}, \citenamefont {Fang},\ and\ \citenamefont
  {Ni}}]{Zhang2022aot}%
  \BibitemOpen
  \bibfield  {author} {\bibinfo {author} {\bibfnamefont {J.~T.}\ \bibnamefont
  {Zhang}}, \bibinfo {author} {\bibfnamefont {L.~R.~B.}\ \bibnamefont
  {Picard}}, \bibinfo {author} {\bibfnamefont {W.~B.}\ \bibnamefont
  {Cairncross}}, \bibinfo {author} {\bibfnamefont {K.}~\bibnamefont {Wang}},
  \bibinfo {author} {\bibfnamefont {Y.}~\bibnamefont {Yu}}, \bibinfo {author}
  {\bibfnamefont {F.}~\bibnamefont {Fang}},\ and\ \bibinfo {author}
  {\bibfnamefont {K.-K.}\ \bibnamefont {Ni}},\ }\bibfield  {title} {\bibinfo
  {title} {An optical tweezer array of ground-state polar molecules},\ }\href
  {https://doi.org/10.1088/2058-9565/ac676c} {\bibfield  {journal} {\bibinfo
  {journal} {Quantum Science and Technology}\ }\textbf {\bibinfo {volume}
  {7}},\ \bibinfo {pages} {035006} (\bibinfo {year} {2022})}\BibitemShut
  {NoStop}%
\bibitem [{\citenamefont {Holland}\ \emph {et~al.}(2023)\citenamefont
  {Holland}, \citenamefont {Lu},\ and\ \citenamefont {Cheuk}}]{Holland2023bio}%
  \BibitemOpen
  \bibfield  {author} {\bibinfo {author} {\bibfnamefont {C.~M.}\ \bibnamefont
  {Holland}}, \bibinfo {author} {\bibfnamefont {Y.}~\bibnamefont {Lu}},\ and\
  \bibinfo {author} {\bibfnamefont {L.~W.}\ \bibnamefont {Cheuk}},\ }\bibfield
  {title} {\bibinfo {title} {Bichromatic imaging of single molecules in an
  optical tweezer array},\ }\href
  {https://doi.org/10.1103/PhysRevLett.131.053202} {\bibfield  {journal}
  {\bibinfo  {journal} {Phys. Rev. Lett.}\ }\textbf {\bibinfo {volume} {131}},\
  \bibinfo {pages} {053202} (\bibinfo {year} {2023})}\BibitemShut {NoStop}%
\bibitem [{\citenamefont {Vilas}\ \emph {et~al.}(2024)\citenamefont {Vilas},
  \citenamefont {Robichaud}, \citenamefont {Hallas}, \citenamefont {Li},
  \citenamefont {Anderegg},\ and\ \citenamefont {Doyle}}]{Vilas2024aot}%
  \BibitemOpen
  \bibfield  {author} {\bibinfo {author} {\bibfnamefont {N.~B.}\ \bibnamefont
  {Vilas}}, \bibinfo {author} {\bibfnamefont {P.}~\bibnamefont {Robichaud}},
  \bibinfo {author} {\bibfnamefont {C.}~\bibnamefont {Hallas}}, \bibinfo
  {author} {\bibfnamefont {G.~K.}\ \bibnamefont {Li}}, \bibinfo {author}
  {\bibfnamefont {L.}~\bibnamefont {Anderegg}},\ and\ \bibinfo {author}
  {\bibfnamefont {J.~M.}\ \bibnamefont {Doyle}},\ }\bibfield  {title} {\bibinfo
  {title} {An optical tweezer array of ultracold polyatomic molecules},\ }\href
  {https://doi.org/10.1038/s41586-024-07199-1} {\bibfield  {journal} {\bibinfo
  {journal} {Nature}\ }\textbf {\bibinfo {volume} {628}},\ \bibinfo {pages}
  {282} (\bibinfo {year} {2024})}\BibitemShut {NoStop}%
\bibitem [{\citenamefont {Ban}\ \emph {et~al.}(2005)\citenamefont {Ban},
  \citenamefont {Jacka}, \citenamefont {Hanssen}, \citenamefont {Reader},\ and\
  \citenamefont {McClelland}}]{Ban2005lct}%
  \BibitemOpen
  \bibfield  {author} {\bibinfo {author} {\bibfnamefont {H.~Y.}\ \bibnamefont
  {Ban}}, \bibinfo {author} {\bibfnamefont {M.}~\bibnamefont {Jacka}}, \bibinfo
  {author} {\bibfnamefont {J.~L.}\ \bibnamefont {Hanssen}}, \bibinfo {author}
  {\bibfnamefont {J.}~\bibnamefont {Reader}},\ and\ \bibinfo {author}
  {\bibfnamefont {J.~J.}\ \bibnamefont {McClelland}},\ }\bibfield  {title}
  {\bibinfo {title} {{Laser cooling transitions in atomic erbium}},\ }\href
  {https://doi.org/10.1364/OPEX.13.003185} {\bibfield  {journal} {\bibinfo
  {journal} {Opt. Express}\ }\textbf {\bibinfo {volume} {13}},\ \bibinfo
  {pages} {3185} (\bibinfo {year} {2005})}\BibitemShut {NoStop}%
\bibitem [{\citenamefont {Picard}\ \emph {et~al.}(2019)\citenamefont {Picard},
  \citenamefont {Mark}, \citenamefont {Ferlaino},\ and\ \citenamefont {van
  Bijnen}}]{Picard2019dla}%
  \BibitemOpen
  \bibfield  {author} {\bibinfo {author} {\bibfnamefont {L.~R.~B.}\
  \bibnamefont {Picard}}, \bibinfo {author} {\bibfnamefont {M.~J.}\
  \bibnamefont {Mark}}, \bibinfo {author} {\bibfnamefont {F.}~\bibnamefont
  {Ferlaino}},\ and\ \bibinfo {author} {\bibfnamefont {R.}~\bibnamefont {van
  Bijnen}},\ }\bibfield  {title} {\bibinfo {title} {Deep learning-assisted
  classification of site-resolved quantum gas microscope images},\ }\href
  {https://doi.org/10.1088/1361-6501/ab44d8} {\bibfield  {journal} {\bibinfo
  {journal} {Measurement Science and Technology}\ }\textbf {\bibinfo {volume}
  {31}},\ \bibinfo {pages} {025201} (\bibinfo {year} {2019})}\BibitemShut
  {NoStop}%
\bibitem [{\citenamefont {Su}\ \emph {et~al.}(2024)\citenamefont {Su},
  \citenamefont {Douglas}, \citenamefont {Szurek}, \citenamefont {Hebert},
  \citenamefont {Krahn}, \citenamefont {Groth}, \citenamefont {Phelps},
  \citenamefont {Markovic},\ and\ \citenamefont {Greiner}}]{Su2024fsa}%
  \BibitemOpen
  \bibfield  {author} {\bibinfo {author} {\bibfnamefont {L.}~\bibnamefont
  {Su}}, \bibinfo {author} {\bibfnamefont {A.}~\bibnamefont {Douglas}},
  \bibinfo {author} {\bibfnamefont {M.}~\bibnamefont {Szurek}}, \bibinfo
  {author} {\bibfnamefont {A.~H.}\ \bibnamefont {Hebert}}, \bibinfo {author}
  {\bibfnamefont {A.}~\bibnamefont {Krahn}}, \bibinfo {author} {\bibfnamefont
  {R.}~\bibnamefont {Groth}}, \bibinfo {author} {\bibfnamefont {G.~A.}\
  \bibnamefont {Phelps}}, \bibinfo {author} {\bibfnamefont {O.}~\bibnamefont
  {Markovic}},\ and\ \bibinfo {author} {\bibfnamefont {M.}~\bibnamefont
  {Greiner}},\ }\bibfield  {title} {\bibinfo {title} {Fast single atom imaging
  in optical lattice arrays},\ }\href@noop {} {\  (\bibinfo {year} {2024})},\
  \Eprint {https://arxiv.org/abs/2404.09978} {arXiv:2404.09978
  [cond-mat.quant-gas]} \BibitemShut {NoStop}%
\bibitem [{\citenamefont {Trautmann}\ \emph {et~al.}(2021)\citenamefont
  {Trautmann}, \citenamefont {Mark}, \citenamefont {Ilzh\"ofer}, \citenamefont
  {Edri}, \citenamefont {Arrach}, \citenamefont {Maloberti}, \citenamefont
  {Greene}, \citenamefont {Robicheaux},\ and\ \citenamefont
  {Ferlaino}}]{Trautmann2021sor}%
  \BibitemOpen
  \bibfield  {author} {\bibinfo {author} {\bibfnamefont {A.}~\bibnamefont
  {Trautmann}}, \bibinfo {author} {\bibfnamefont {M.~J.}\ \bibnamefont {Mark}},
  \bibinfo {author} {\bibfnamefont {P.}~\bibnamefont {Ilzh\"ofer}}, \bibinfo
  {author} {\bibfnamefont {H.}~\bibnamefont {Edri}}, \bibinfo {author}
  {\bibfnamefont {A.~E.}\ \bibnamefont {Arrach}}, \bibinfo {author}
  {\bibfnamefont {J.~G.}\ \bibnamefont {Maloberti}}, \bibinfo {author}
  {\bibfnamefont {C.~H.}\ \bibnamefont {Greene}}, \bibinfo {author}
  {\bibfnamefont {F.}~\bibnamefont {Robicheaux}},\ and\ \bibinfo {author}
  {\bibfnamefont {F.}~\bibnamefont {Ferlaino}},\ }\bibfield  {title} {\bibinfo
  {title} {{Spectroscopy of Rydberg states in erbium using electromagnetically
  induced transparency}},\ }\href
  {https://doi.org/10.1103/PhysRevResearch.3.033165} {\bibfield  {journal}
  {\bibinfo  {journal} {Phys. Rev. Res.}\ }\textbf {\bibinfo {volume} {3}},\
  \bibinfo {pages} {033165} (\bibinfo {year} {2021})}\BibitemShut {NoStop}%
\bibitem [{\citenamefont {Kruckenhauser}\ \emph {et~al.}(2022)\citenamefont
  {Kruckenhauser}, \citenamefont {van Bijnen}, \citenamefont {Zache},
  \citenamefont {Liberto},\ and\ \citenamefont
  {Zoller}}]{Kruckenhauser2023hds}%
  \BibitemOpen
  \bibfield  {author} {\bibinfo {author} {\bibfnamefont {A.}~\bibnamefont
  {Kruckenhauser}}, \bibinfo {author} {\bibfnamefont {R.}~\bibnamefont {van
  Bijnen}}, \bibinfo {author} {\bibfnamefont {T.~V.}\ \bibnamefont {Zache}},
  \bibinfo {author} {\bibfnamefont {M.~D.}\ \bibnamefont {Liberto}},\ and\
  \bibinfo {author} {\bibfnamefont {P.}~\bibnamefont {Zoller}},\ }\bibfield
  {title} {\bibinfo {title} {High-dimensional so(4)-symmetric rydberg manifolds
  for quantum simulation},\ }\href {https://doi.org/10.1088/2058-9565/aca996}
  {\bibfield  {journal} {\bibinfo  {journal} {Quantum Science and Technology}\
  }\textbf {\bibinfo {volume} {8}},\ \bibinfo {pages} {015020} (\bibinfo {year}
  {2022})}\BibitemShut {NoStop}%
\bibitem [{\citenamefont {Norcia}\ and\ \citenamefont
  {Ferlaino}(2021)}]{Norcia2021dia}%
  \BibitemOpen
  \bibfield  {author} {\bibinfo {author} {\bibfnamefont {M.~A.}\ \bibnamefont
  {Norcia}}\ and\ \bibinfo {author} {\bibfnamefont {F.}~\bibnamefont
  {Ferlaino}},\ }\bibfield  {title} {\bibinfo {title} {{Developments in atomic
  control using ultracold magnetic lanthanides}},\ }\href
  {https://doi.org/10.1038/s41567-021-01398-7} {\bibfield  {journal} {\bibinfo
  {journal} {Nature Physics}\ }\textbf {\bibinfo {volume} {17}},\ \bibinfo
  {pages} {1349} (\bibinfo {year} {2021})}\BibitemShut {NoStop}%
\bibitem [{\citenamefont {Dzuba}\ \emph {et~al.}(2011)\citenamefont {Dzuba},
  \citenamefont {Flambaum},\ and\ \citenamefont {Lev}}]{Dzuba2011dpa}%
  \BibitemOpen
  \bibfield  {author} {\bibinfo {author} {\bibfnamefont {V.~A.}\ \bibnamefont
  {Dzuba}}, \bibinfo {author} {\bibfnamefont {V.~V.}\ \bibnamefont
  {Flambaum}},\ and\ \bibinfo {author} {\bibfnamefont {B.~L.}\ \bibnamefont
  {Lev}},\ }\bibfield  {title} {\bibinfo {title} {{Dynamic polarizabilities and
  magic wavelengths for dysprosium}},\ }\href
  {https://doi.org/10.1103/PhysRevA.83.032502} {\bibfield  {journal} {\bibinfo
  {journal} {Phys. Rev. A}\ }\textbf {\bibinfo {volume} {83}},\ \bibinfo
  {pages} {032502} (\bibinfo {year} {2011})}\BibitemShut {NoStop}%
\bibitem [{\citenamefont {Becher}\ \emph {et~al.}(2018)\citenamefont {Becher},
  \citenamefont {Baier}, \citenamefont {Aikawa}, \citenamefont {Lepers},
  \citenamefont {Wyart}, \citenamefont {Dulieu},\ and\ \citenamefont
  {Ferlaino}}]{Becher2018apo}%
  \BibitemOpen
  \bibfield  {author} {\bibinfo {author} {\bibfnamefont {J.~H.}\ \bibnamefont
  {Becher}}, \bibinfo {author} {\bibfnamefont {S.}~\bibnamefont {Baier}},
  \bibinfo {author} {\bibfnamefont {K.}~\bibnamefont {Aikawa}}, \bibinfo
  {author} {\bibfnamefont {M.}~\bibnamefont {Lepers}}, \bibinfo {author}
  {\bibfnamefont {J.-F.}\ \bibnamefont {Wyart}}, \bibinfo {author}
  {\bibfnamefont {O.}~\bibnamefont {Dulieu}},\ and\ \bibinfo {author}
  {\bibfnamefont {F.}~\bibnamefont {Ferlaino}},\ }\bibfield  {title} {\bibinfo
  {title} {{Anisotropic polarizability of erbium atoms}},\ }\href
  {https://doi.org/10.1103/PhysRevA.97.012509} {\bibfield  {journal} {\bibinfo
  {journal} {Phys. Rev. A}\ }\textbf {\bibinfo {volume} {97}},\ \bibinfo
  {pages} {012509} (\bibinfo {year} {2018})}\BibitemShut {NoStop}%
\bibitem [{\citenamefont {Chalopin}\ \emph {et~al.}(2018)\citenamefont
  {Chalopin}, \citenamefont {Makhalov}, \citenamefont {Bouazza}, \citenamefont
  {Evrard}, \citenamefont {Barker}, \citenamefont {Lepers}, \citenamefont
  {Wyart}, \citenamefont {Dulieu}, \citenamefont {Dalibard}, \citenamefont
  {Lopes},\ and\ \citenamefont {Nascimbene}}]{Chalopin2018als}%
  \BibitemOpen
  \bibfield  {author} {\bibinfo {author} {\bibfnamefont {T.}~\bibnamefont
  {Chalopin}}, \bibinfo {author} {\bibfnamefont {V.}~\bibnamefont {Makhalov}},
  \bibinfo {author} {\bibfnamefont {C.}~\bibnamefont {Bouazza}}, \bibinfo
  {author} {\bibfnamefont {A.}~\bibnamefont {Evrard}}, \bibinfo {author}
  {\bibfnamefont {A.}~\bibnamefont {Barker}}, \bibinfo {author} {\bibfnamefont
  {M.}~\bibnamefont {Lepers}}, \bibinfo {author} {\bibfnamefont {J.-F.}\
  \bibnamefont {Wyart}}, \bibinfo {author} {\bibfnamefont {O.}~\bibnamefont
  {Dulieu}}, \bibinfo {author} {\bibfnamefont {J.}~\bibnamefont {Dalibard}},
  \bibinfo {author} {\bibfnamefont {R.}~\bibnamefont {Lopes}},\ and\ \bibinfo
  {author} {\bibfnamefont {S.}~\bibnamefont {Nascimbene}},\ }\bibfield  {title}
  {\bibinfo {title} {{Anisotropic light shift and magic polarization of the
  intercombination line of dysprosium atoms in a far-detuned dipole trap}},\
  }\href {https://doi.org/10.1103/PhysRevA.98.040502} {\bibfield  {journal}
  {\bibinfo  {journal} {Phys. Rev. A}\ }\textbf {\bibinfo {volume} {98}},\
  \bibinfo {pages} {040502(R)} (\bibinfo {year} {2018})}\BibitemShut {NoStop}%
\bibitem [{\citenamefont {Bloch}\ \emph {et~al.}(2024)\citenamefont {Bloch},
  \citenamefont {Hofer}, \citenamefont {Cohen}, \citenamefont {Lepers},
  \citenamefont {Browaeys},\ and\ \citenamefont
  {Ferrier-Barbut}}]{Bloch2024apo}%
  \BibitemOpen
  \bibfield  {author} {\bibinfo {author} {\bibfnamefont {D.}~\bibnamefont
  {Bloch}}, \bibinfo {author} {\bibfnamefont {B.}~\bibnamefont {Hofer}},
  \bibinfo {author} {\bibfnamefont {S.~R.}\ \bibnamefont {Cohen}}, \bibinfo
  {author} {\bibfnamefont {M.}~\bibnamefont {Lepers}}, \bibinfo {author}
  {\bibfnamefont {A.}~\bibnamefont {Browaeys}},\ and\ \bibinfo {author}
  {\bibfnamefont {I.}~\bibnamefont {Ferrier-Barbut}},\ }\href@noop {} {\bibinfo
  {title} {Anisotropic polarizability of {D}y at 532 nm on the intercombination
  transition}} (\bibinfo {year} {2024}),\ \Eprint
  {https://arxiv.org/abs/2404.10480} {arXiv:2404.10480 [physics.atom-ph]}
  \BibitemShut {NoStop}%
\bibitem [{\citenamefont {Bloch}\ \emph {et~al.}(2023)\citenamefont {Bloch},
  \citenamefont {Hofer}, \citenamefont {Cohen}, \citenamefont {Browaeys},\ and\
  \citenamefont {Ferrier-Barbut}}]{Bloch2023tai}%
  \BibitemOpen
  \bibfield  {author} {\bibinfo {author} {\bibfnamefont {D.}~\bibnamefont
  {Bloch}}, \bibinfo {author} {\bibfnamefont {B.}~\bibnamefont {Hofer}},
  \bibinfo {author} {\bibfnamefont {S.~R.}\ \bibnamefont {Cohen}}, \bibinfo
  {author} {\bibfnamefont {A.}~\bibnamefont {Browaeys}},\ and\ \bibinfo
  {author} {\bibfnamefont {I.}~\bibnamefont {Ferrier-Barbut}},\ }\bibfield
  {title} {\bibinfo {title} {Trapping and imaging single dysprosium atoms in
  optical tweezer arrays},\ }\href
  {https://doi.org/10.1103/PhysRevLett.131.203401} {\bibfield  {journal}
  {\bibinfo  {journal} {Phys. Rev. Lett.}\ }\textbf {\bibinfo {volume} {131}},\
  \bibinfo {pages} {203401} (\bibinfo {year} {2023})}\BibitemShut {NoStop}%
\bibitem [{\citenamefont {Frisch}\ \emph {et~al.}(2012)\citenamefont {Frisch},
  \citenamefont {Aikawa}, \citenamefont {Mark}, \citenamefont {Rietzler},
  \citenamefont {Schindler}, \citenamefont {Zupani\ifmmode~\check{c}\else
  \v{c}\fi{}}, \citenamefont {Grimm},\ and\ \citenamefont
  {Ferlaino}}]{Frisch2012narrowline}%
  \BibitemOpen
  \bibfield  {author} {\bibinfo {author} {\bibfnamefont {A.}~\bibnamefont
  {Frisch}}, \bibinfo {author} {\bibfnamefont {K.}~\bibnamefont {Aikawa}},
  \bibinfo {author} {\bibfnamefont {M.}~\bibnamefont {Mark}}, \bibinfo {author}
  {\bibfnamefont {A.}~\bibnamefont {Rietzler}}, \bibinfo {author}
  {\bibfnamefont {J.}~\bibnamefont {Schindler}}, \bibinfo {author}
  {\bibfnamefont {E.}~\bibnamefont {Zupani\ifmmode~\check{c}\else \v{c}\fi{}}},
  \bibinfo {author} {\bibfnamefont {R.}~\bibnamefont {Grimm}},\ and\ \bibinfo
  {author} {\bibfnamefont {F.}~\bibnamefont {Ferlaino}},\ }\bibfield  {title}
  {\bibinfo {title} {Narrow-line magneto-optical trap for erbium},\ }\href
  {https://doi.org/10.1103/PhysRevA.85.051401} {\bibfield  {journal} {\bibinfo
  {journal} {Phys. Rev. A}\ }\textbf {\bibinfo {volume} {85}},\ \bibinfo
  {pages} {051401} (\bibinfo {year} {2012})}\BibitemShut {NoStop}%
\bibitem [{\citenamefont {Madjarov}\ \emph {et~al.}(2020)\citenamefont
  {Madjarov}, \citenamefont {Covey}, \citenamefont {Shaw}, \citenamefont
  {Choi}, \citenamefont {Kale}, \citenamefont {Cooper}, \citenamefont
  {Pichler}, \citenamefont {Schkolnik}, \citenamefont {Williams},\ and\
  \citenamefont {Endres}}]{Madjarov2020hfe}%
  \BibitemOpen
  \bibfield  {author} {\bibinfo {author} {\bibfnamefont {I.~S.}\ \bibnamefont
  {Madjarov}}, \bibinfo {author} {\bibfnamefont {J.~P.}\ \bibnamefont {Covey}},
  \bibinfo {author} {\bibfnamefont {A.~L.}\ \bibnamefont {Shaw}}, \bibinfo
  {author} {\bibfnamefont {J.}~\bibnamefont {Choi}}, \bibinfo {author}
  {\bibfnamefont {A.}~\bibnamefont {Kale}}, \bibinfo {author} {\bibfnamefont
  {A.}~\bibnamefont {Cooper}}, \bibinfo {author} {\bibfnamefont
  {H.}~\bibnamefont {Pichler}}, \bibinfo {author} {\bibfnamefont
  {V.}~\bibnamefont {Schkolnik}}, \bibinfo {author} {\bibfnamefont {J.~R.}\
  \bibnamefont {Williams}},\ and\ \bibinfo {author} {\bibfnamefont
  {M.}~\bibnamefont {Endres}},\ }\bibfield  {title} {\bibinfo {title}
  {High-fidelity entanglement and detection of alkaline-earth {R}ydberg
  atoms},\ }\href {https://doi.org/10.1038/s41567-020-0903-z} {\bibfield
  {journal} {\bibinfo  {journal} {Nature Physics}\ }\textbf {\bibinfo {volume}
  {16}},\ \bibinfo {pages} {857} (\bibinfo {year} {2020})}\BibitemShut
  {NoStop}%
\bibitem [{\citenamefont {Patscheider}\ \emph {et~al.}(2021)\citenamefont
  {Patscheider}, \citenamefont {Yang}, \citenamefont {Natale}, \citenamefont
  {Petter}, \citenamefont {Chomaz}, \citenamefont {Mark}, \citenamefont
  {Hovhannesyan}, \citenamefont {Lepers},\ and\ \citenamefont
  {Ferlaino}}]{Patscheider2021ooa}%
  \BibitemOpen
  \bibfield  {author} {\bibinfo {author} {\bibfnamefont {A.}~\bibnamefont
  {Patscheider}}, \bibinfo {author} {\bibfnamefont {B.}~\bibnamefont {Yang}},
  \bibinfo {author} {\bibfnamefont {G.}~\bibnamefont {Natale}}, \bibinfo
  {author} {\bibfnamefont {D.}~\bibnamefont {Petter}}, \bibinfo {author}
  {\bibfnamefont {L.}~\bibnamefont {Chomaz}}, \bibinfo {author} {\bibfnamefont
  {M.~J.}\ \bibnamefont {Mark}}, \bibinfo {author} {\bibfnamefont
  {G.}~\bibnamefont {Hovhannesyan}}, \bibinfo {author} {\bibfnamefont
  {M.}~\bibnamefont {Lepers}},\ and\ \bibinfo {author} {\bibfnamefont
  {F.}~\bibnamefont {Ferlaino}},\ }\bibfield  {title} {\bibinfo {title}
  {{Observation of a narrow inner-shell orbital transition in atomic erbium at
  1299 nm}},\ }\href {https://doi.org/10.1103/PhysRevResearch.3.033256}
  {\bibfield  {journal} {\bibinfo  {journal} {Phys. Rev. Research}\ }\textbf
  {\bibinfo {volume} {3}},\ \bibinfo {pages} {033256} (\bibinfo {year}
  {2021})}\BibitemShut {NoStop}%
\bibitem [{\citenamefont {Ilzh\"ofer}\ \emph {et~al.}(2018)\citenamefont
  {Ilzh\"ofer}, \citenamefont {Durastante}, \citenamefont {Patscheider},
  \citenamefont {Trautmann}, \citenamefont {Mark},\ and\ \citenamefont
  {Ferlaino}}]{Ilzhoefer2018tsf}%
  \BibitemOpen
  \bibfield  {author} {\bibinfo {author} {\bibfnamefont {P.}~\bibnamefont
  {Ilzh\"ofer}}, \bibinfo {author} {\bibfnamefont {G.}~\bibnamefont
  {Durastante}}, \bibinfo {author} {\bibfnamefont {A.}~\bibnamefont
  {Patscheider}}, \bibinfo {author} {\bibfnamefont {A.}~\bibnamefont
  {Trautmann}}, \bibinfo {author} {\bibfnamefont {M.~J.}\ \bibnamefont
  {Mark}},\ and\ \bibinfo {author} {\bibfnamefont {F.}~\bibnamefont
  {Ferlaino}},\ }\bibfield  {title} {\bibinfo {title} {{Two-species five-beam
  magneto-optical trap for erbium and dysprosium}},\ }\href
  {https://doi.org/10.1103/PhysRevA.97.023633} {\bibfield  {journal} {\bibinfo
  {journal} {Phys. Rev. A}\ }\textbf {\bibinfo {volume} {97}},\ \bibinfo
  {pages} {023633} (\bibinfo {year} {2018})}\BibitemShut {NoStop}%
\bibitem [{sup()}]{supmat}%
  \BibitemOpen
  \href@noop {} {\emph {\bibinfo {title} {\textnormal{See supplemental
  material}}}}\BibitemShut {NoStop}%
\bibitem [{\citenamefont {Aikawa}\ \emph {et~al.}(2012)\citenamefont {Aikawa},
  \citenamefont {Frisch}, \citenamefont {Mark}, \citenamefont {Baier},
  \citenamefont {Rietzler}, \citenamefont {Grimm},\ and\ \citenamefont
  {Ferlaino}}]{Aikawa:2012}%
  \BibitemOpen
  \bibfield  {author} {\bibinfo {author} {\bibfnamefont {K.}~\bibnamefont
  {Aikawa}}, \bibinfo {author} {\bibfnamefont {A.}~\bibnamefont {Frisch}},
  \bibinfo {author} {\bibfnamefont {M.}~\bibnamefont {Mark}}, \bibinfo {author}
  {\bibfnamefont {S.}~\bibnamefont {Baier}}, \bibinfo {author} {\bibfnamefont
  {A.}~\bibnamefont {Rietzler}}, \bibinfo {author} {\bibfnamefont
  {R.}~\bibnamefont {Grimm}},\ and\ \bibinfo {author} {\bibfnamefont
  {F.}~\bibnamefont {Ferlaino}},\ }\bibfield  {title} {\bibinfo {title}
  {{B}ose-{E}instein condensation of erbium},\ }\href
  {https://doi.org/10.1103/PhysRevLett.108.210401} {\bibfield  {journal}
  {\bibinfo  {journal} {Phys. Rev. Lett.}\ }\textbf {\bibinfo {volume} {108}},\
  \bibinfo {pages} {210401} (\bibinfo {year} {2012})}\BibitemShut {NoStop}%
\bibitem [{\citenamefont {Miranda}\ \emph {et~al.}(2015)\citenamefont
  {Miranda}, \citenamefont {Inoue}, \citenamefont {Okuyama}, \citenamefont
  {Nakamoto},\ and\ \citenamefont {Kozuma}}]{Miranda2015sri}%
  \BibitemOpen
  \bibfield  {author} {\bibinfo {author} {\bibfnamefont {M.}~\bibnamefont
  {Miranda}}, \bibinfo {author} {\bibfnamefont {R.}~\bibnamefont {Inoue}},
  \bibinfo {author} {\bibfnamefont {Y.}~\bibnamefont {Okuyama}}, \bibinfo
  {author} {\bibfnamefont {A.}~\bibnamefont {Nakamoto}},\ and\ \bibinfo
  {author} {\bibfnamefont {M.}~\bibnamefont {Kozuma}},\ }\bibfield  {title}
  {\bibinfo {title} {Site-resolved imaging of ytterbium atoms in a
  two-dimensional optical lattice},\ }\href
  {https://doi.org/10.1103/PhysRevA.91.063414} {\bibfield  {journal} {\bibinfo
  {journal} {Phys. Rev. A}\ }\textbf {\bibinfo {volume} {91}},\ \bibinfo
  {pages} {063414} (\bibinfo {year} {2015})}\BibitemShut {NoStop}%
\bibitem [{\citenamefont {Su}\ \emph {et~al.}(2023)\citenamefont {Su},
  \citenamefont {Douglas}, \citenamefont {Szurek}, \citenamefont {Groth},
  \citenamefont {Ozturk}, \citenamefont {Krahn}, \citenamefont {H{\'e}bert},
  \citenamefont {Phelps}, \citenamefont {Ebadi}, \citenamefont {Dickerson}
  \emph {et~al.}}]{Su2023dqs_}%
  \BibitemOpen
  \bibfield  {author} {\bibinfo {author} {\bibfnamefont {L.}~\bibnamefont
  {Su}}, \bibinfo {author} {\bibfnamefont {A.}~\bibnamefont {Douglas}},
  \bibinfo {author} {\bibfnamefont {M.}~\bibnamefont {Szurek}}, \bibinfo
  {author} {\bibfnamefont {R.}~\bibnamefont {Groth}}, \bibinfo {author}
  {\bibfnamefont {S.~F.}\ \bibnamefont {Ozturk}}, \bibinfo {author}
  {\bibfnamefont {A.}~\bibnamefont {Krahn}}, \bibinfo {author} {\bibfnamefont
  {A.~H.}\ \bibnamefont {H{\'e}bert}}, \bibinfo {author} {\bibfnamefont
  {G.~A.}\ \bibnamefont {Phelps}}, \bibinfo {author} {\bibfnamefont
  {S.}~\bibnamefont {Ebadi}}, \bibinfo {author} {\bibfnamefont
  {S.}~\bibnamefont {Dickerson}}, \emph {et~al.},\ }\bibfield  {title}
  {\bibinfo {title} {Dipolar quantum solids emerging in a {H}ubbard quantum
  simulator},\ }\href {https://doi.org/10.1038/s41586-023-06614-3} {\bibfield
  {journal} {\bibinfo  {journal} {Nature}\ }\textbf {\bibinfo {volume} {622}},\
  \bibinfo {pages} {724} (\bibinfo {year} {2023})}\BibitemShut {NoStop}%
\bibitem [{\citenamefont {Lepers}\ \emph {et~al.}(2014)\citenamefont {Lepers},
  \citenamefont {Wyart},\ and\ \citenamefont {Dulieu}}]{Lepers2014aot}%
  \BibitemOpen
  \bibfield  {author} {\bibinfo {author} {\bibfnamefont {M.}~\bibnamefont
  {Lepers}}, \bibinfo {author} {\bibfnamefont {J.-F.}\ \bibnamefont {Wyart}},\
  and\ \bibinfo {author} {\bibfnamefont {O.}~\bibnamefont {Dulieu}},\
  }\bibfield  {title} {\bibinfo {title} {{Anisotropic optical trapping of
  ultracold erbium atoms}},\ }\href
  {https://doi.org/10.1103/PhysRevA.89.022505} {\bibfield  {journal} {\bibinfo
  {journal} {Phys. Rev. A}\ }\textbf {\bibinfo {volume} {89}},\ \bibinfo
  {pages} {022505} (\bibinfo {year} {2014})}\BibitemShut {NoStop}%
\bibitem [{\citenamefont {Rosenband}\ \emph {et~al.}(2018)\citenamefont
  {Rosenband}, \citenamefont {Grimes},\ and\ \citenamefont
  {Ni}}]{Rosenband2018sel}%
  \BibitemOpen
  \bibfield  {author} {\bibinfo {author} {\bibfnamefont {T.}~\bibnamefont
  {Rosenband}}, \bibinfo {author} {\bibfnamefont {D.~D.}\ \bibnamefont
  {Grimes}},\ and\ \bibinfo {author} {\bibfnamefont {K.-K.}\ \bibnamefont
  {Ni}},\ }\bibfield  {title} {\bibinfo {title} {Elliptical polarization for
  molecular {S}tark shift compensation in deep optical traps},\ }\href
  {https://doi.org/10.1364/OE.26.019821} {\bibfield  {journal} {\bibinfo
  {journal} {Opt. Express}\ }\textbf {\bibinfo {volume} {26}},\ \bibinfo
  {pages} {19821} (\bibinfo {year} {2018})}\BibitemShut {NoStop}%
\bibitem [{\citenamefont {Thompson}\ \emph {et~al.}(2013)\citenamefont
  {Thompson}, \citenamefont {Tiecke}, \citenamefont {Zibrov}, \citenamefont
  {Vuleti\ifmmode~\acute{c}\else \'{c}\fi{}},\ and\ \citenamefont
  {Lukin}}]{Thompson2013car}%
  \BibitemOpen
  \bibfield  {author} {\bibinfo {author} {\bibfnamefont {J.~D.}\ \bibnamefont
  {Thompson}}, \bibinfo {author} {\bibfnamefont {T.~G.}\ \bibnamefont
  {Tiecke}}, \bibinfo {author} {\bibfnamefont {A.~S.}\ \bibnamefont {Zibrov}},
  \bibinfo {author} {\bibfnamefont {V.}~\bibnamefont
  {Vuleti\ifmmode~\acute{c}\else \'{c}\fi{}}},\ and\ \bibinfo {author}
  {\bibfnamefont {M.~D.}\ \bibnamefont {Lukin}},\ }\bibfield  {title} {\bibinfo
  {title} {Coherence and {R}aman sideband cooling of a single atom in an
  optical tweezer},\ }\href {https://doi.org/10.1103/PhysRevLett.110.133001}
  {\bibfield  {journal} {\bibinfo  {journal} {Phys. Rev. Lett.}\ }\textbf
  {\bibinfo {volume} {110}},\ \bibinfo {pages} {133001} (\bibinfo {year}
  {2013})}\BibitemShut {NoStop}%
\bibitem [{\citenamefont {Weiner}\ \emph {et~al.}(1999)\citenamefont {Weiner},
  \citenamefont {Bagnato}, \citenamefont {Zilio},\ and\ \citenamefont
  {Julienne}}]{Weiner1999eat}%
  \BibitemOpen
  \bibfield  {author} {\bibinfo {author} {\bibfnamefont {J.}~\bibnamefont
  {Weiner}}, \bibinfo {author} {\bibfnamefont {V.~S.}\ \bibnamefont {Bagnato}},
  \bibinfo {author} {\bibfnamefont {S.}~\bibnamefont {Zilio}},\ and\ \bibinfo
  {author} {\bibfnamefont {P.~S.}\ \bibnamefont {Julienne}},\ }\bibfield
  {title} {\bibinfo {title} {Experiments and theory in cold and ultracold
  collisions},\ }\href {https://doi.org/10.1103/RevModPhys.71.1} {\bibfield
  {journal} {\bibinfo  {journal} {Rev. Mod. Phys.}\ }\textbf {\bibinfo {volume}
  {71}},\ \bibinfo {pages} {1} (\bibinfo {year} {1999})}\BibitemShut {NoStop}%
\bibitem [{\citenamefont {Schlosser}\ \emph {et~al.}(2002)\citenamefont
  {Schlosser}, \citenamefont {Reymond},\ and\ \citenamefont
  {Grangier}}]{Schlosser2002cbi}%
  \BibitemOpen
  \bibfield  {author} {\bibinfo {author} {\bibfnamefont {N.}~\bibnamefont
  {Schlosser}}, \bibinfo {author} {\bibfnamefont {G.}~\bibnamefont {Reymond}},\
  and\ \bibinfo {author} {\bibfnamefont {P.}~\bibnamefont {Grangier}},\
  }\bibfield  {title} {\bibinfo {title} {Collisional blockade in microscopic
  optical dipole traps},\ }\href
  {https://doi.org/10.1103/PhysRevLett.89.023005} {\bibfield  {journal}
  {\bibinfo  {journal} {Phys. Rev. Lett.}\ }\textbf {\bibinfo {volume} {89}},\
  \bibinfo {pages} {023005} (\bibinfo {year} {2002})}\BibitemShut {NoStop}%
\bibitem [{\citenamefont {Fung}\ \emph {et~al.}(2016)\citenamefont {Fung},
  \citenamefont {Sompet},\ and\ \citenamefont {Andersen}}]{Fung2016sap}%
  \BibitemOpen
  \bibfield  {author} {\bibinfo {author} {\bibfnamefont {Y.~H.}\ \bibnamefont
  {Fung}}, \bibinfo {author} {\bibfnamefont {P.}~\bibnamefont {Sompet}},\ and\
  \bibinfo {author} {\bibfnamefont {M.~F.}\ \bibnamefont {Andersen}},\
  }\bibfield  {title} {\bibinfo {title} {Single atoms preparation using
  light-assisted collisions},\ }\bibfield  {journal} {\bibinfo  {journal}
  {Technologies}\ }\textbf {\bibinfo {volume} {4}},\ \href
  {https://doi.org/10.3390/technologies4010004} {10.3390/technologies4010004}
  (\bibinfo {year} {2016})\BibitemShut {NoStop}%
\bibitem [{\citenamefont {Sortais}\ \emph {et~al.}(2012)\citenamefont
  {Sortais}, \citenamefont {Fuhrmanek}, \citenamefont {Bourgain},\ and\
  \citenamefont {Browaeys}}]{Sortais2012spa}%
  \BibitemOpen
  \bibfield  {author} {\bibinfo {author} {\bibfnamefont {Y.~R.~P.}\
  \bibnamefont {Sortais}}, \bibinfo {author} {\bibfnamefont {A.}~\bibnamefont
  {Fuhrmanek}}, \bibinfo {author} {\bibfnamefont {R.}~\bibnamefont
  {Bourgain}},\ and\ \bibinfo {author} {\bibfnamefont {A.}~\bibnamefont
  {Browaeys}},\ }\bibfield  {title} {\bibinfo {title} {Sub-{P}oissonian
  atom-number fluctuations using light-assisted collisions},\ }\href
  {https://doi.org/10.1103/PhysRevA.85.035403} {\bibfield  {journal} {\bibinfo
  {journal} {Phys. Rev. A}\ }\textbf {\bibinfo {volume} {85}},\ \bibinfo
  {pages} {035403} (\bibinfo {year} {2012})}\BibitemShut {NoStop}%
\bibitem [{\citenamefont {Gallagher}\ and\ \citenamefont
  {Pritchard}(1989)}]{Gallagher1989eco}%
  \BibitemOpen
  \bibfield  {author} {\bibinfo {author} {\bibfnamefont {A.}~\bibnamefont
  {Gallagher}}\ and\ \bibinfo {author} {\bibfnamefont {D.~E.}\ \bibnamefont
  {Pritchard}},\ }\bibfield  {title} {\bibinfo {title} {Exoergic collisions of
  cold na*-na},\ }\href {https://doi.org/10.1103/PhysRevLett.63.957} {\bibfield
   {journal} {\bibinfo  {journal} {Phys. Rev. Lett.}\ }\textbf {\bibinfo
  {volume} {63}},\ \bibinfo {pages} {957} (\bibinfo {year} {1989})}\BibitemShut
  {NoStop}%
\bibitem [{\citenamefont {Li}\ \emph {et~al.}(2017)\citenamefont {Li},
  \citenamefont {Wyart}, \citenamefont {Dulieu},\ and\ \citenamefont
  {Lepers}}]{Li2017aot}%
  \BibitemOpen
  \bibfield  {author} {\bibinfo {author} {\bibfnamefont {H.}~\bibnamefont
  {Li}}, \bibinfo {author} {\bibfnamefont {J.-F.}\ \bibnamefont {Wyart}},
  \bibinfo {author} {\bibfnamefont {O.}~\bibnamefont {Dulieu}},\ and\ \bibinfo
  {author} {\bibfnamefont {M.}~\bibnamefont {Lepers}},\ }\bibfield  {title}
  {\bibinfo {title} {{Anisotropic optical trapping as a manifestation of the
  complex electronic structure of ultracold lanthanide atoms: The example of
  holmium}},\ }\href {https://doi.org/10.1103/PhysRevA.95.062508} {\bibfield
  {journal} {\bibinfo  {journal} {Phys. Rev. A}\ }\textbf {\bibinfo {volume}
  {95}},\ \bibinfo {pages} {062508} (\bibinfo {year} {2017})}\BibitemShut
  {NoStop}%
\end{thebibliography}%
